\documentstyle[aps,prl,epsf,amsfonts,amssymb,amsmath]{revtex}

\begin{document}
\advance\textheight by 0.2in
\draft
\twocolumn[\hsize\textwidth\columnwidth\hsize\csname@twocolumnfalse%
\endcsname
\title{Patterns in randomly evolving networks: Idiotypic networks}
\author{M. Brede and U. Behn}
\address{Institut f\"ur Theoretische Physik, Universit\"at Leipzig, 
Augustusplatz 10, 
 D-04109 Leipzig, Germany}
\date{Draft, \today}
\maketitle

\begin{abstract}
We present a model for the evolution of networks of occupied sites on undirected regular graphs. At every iteration step in a parallel update $I$ randomly chosen empty sites are occupied and occupied sites having degree outside of a given interval $(t_{\text{l}},t_{\text{u}})$ are set empty. Depending on the influx $I$ and the values of both lower threshold and upper threshold of the degree different kinds of behaviour can be observed. In certain regimes stable long-living patterns appear. We distinguish two types of pattern: static patterns arising on graphs with low connectivity and dynamic patterns found on high connectivity graphs. Increasing $I$ patterns become unstable and transitions between almost stable patterns, interrupted by disordered phases, occur. For still larger $I$ the lifetime of occupied sites becomes very small and network structures are dominated by randomness. We develop methods to analyze nature and dynamics of these network patterns, give a statistical description of defects and fluctuations around them, and elucidate transitions between different patterns. Results and methods presented can be applied to a variety of problems in different fields and a broad class of graphs. Aiming chiefly at the modeling of functional networks of interacting antibodies and B-cells of the immune system (idiotypic networks) we focus on a class of graphs constructed by bit-chains. The biological relevance of the patterns and possible operational modes of idiotypic networks are discussed.
\end{abstract}

] 
\vspace*{0.5cm}

\section{Introduction}
Many complex coupled systems can be modeled as networks: Vertices being associated with the systems elements and edges representing interactions between them. Metabolic networks \cite{Jeong,FellWagner}, food webs \cite{MontoyaSole}, social networks \cite {Liljeros}, and networks in the immune system (for some approaches see \cite{Ueb}) are just examples of such systems. Frequently, although the detailed interactions of the elements may be rather complicated, the knowledge about the underlying network structure facilitates the understanding of essential features of the system as a whole \cite{Strogatz}.

In the absence of detailed knowledge about a networks topology previous works tended to assume either a completely random link structure \cite{Erdos,Bollobas,Newman} or clusters given by percolation on lattices \cite{Stauffer}. The complexity of many systems, however, emerges as a consequence of some underlying principle for their evolution, e.g., preferential attachment \cite{Barabasi1}, optimization of transportation and communication pathways \cite{West,Mathias,Cancho} extinction of the least populated species \cite{Jain1,Jain2}, or of the mere fact that the network has randomly evolved in time \cite{Callaway}. This, altogether may lead to a highly organized network topology.

Recently, much attention has been devoted to a variety of networks which exhibit topological properties different from random graphs (for a review see \cite{Barabasi2,Wattsbook,Dov}). For obtaining a detailed understanding of their architecture it has proved useful to trace the dynamics of the networks structural growth. As recently observed major transitions in a systems dynamics can even be completely governed by preceding transformations in its network structure \cite{Jain1,Jain2}.

In the case of idiotypic networks (INW's) in the immune system \cite{Jerne}, dynamics and network evolution are driven by a continuous influx of new idiotypes from the bone marrow. Data from experimental investigations suggest a daily bone marrow production only one order smaller than the actual network size \cite{Ueb,Shannon,MacLennan}. A second principle governing the dynamics of INW's is a set of local rules: idiotypes die out if they are under- or overstimulated, i.e. have too few or too many neighbours. As Conways famous `game of life' \cite{Conway} suggests, systems governed by such local rules can evolve towards highly complex self-organized states \cite{Bak}.

Starting from a simple set of rules intending to mimic the dynamics within idiotypic networks in the immune system, we present a cellular automaton based model for the evolution of networks of occupied sites on graphs. Depending on the major parameter, the bone marrow influx, the model exhibits parameter regimes in which the system self-organizes into static and dynamic patterns, shows transitions between such ordered phases and has a parameter range in which it is governed by randomness. We investigate these regimes and elucidate connections between the overall dynamics and the evolution of the network structure.

The organization of this paper is as follows. In the next section we introduce the model, give a brief introduction to idiotypic networks and outline further possible applications of our model. Then, in Sec. \ref{analyzingthedynamics}, we give an overview over the typical evolution of the initial dynamics of the population and its connection with changes in the network structure. Next, in Sec. \ref{thestationarystate} the stationary state and its network organization will be characterized. In the fifth section the dependence of the systems qualitative behaviour on the models main parameter, the influx $I$, will be explored, analytical results for small $I$ derived from a microscopic view of the network characterizing fluctuations around the steady state be presented and an explanation for the systems high $I$ behaviour be given. In the final section changes in the variety of stationary network patterns for more closely linked base graphs will be discussed. 

The aim of this paper is not fine tuning or in detail matching of one of the applications, but rather generally exploring features of the dynamics caused by the window-algorithm and understanding its connection with network structures thus evolving.
\section {The model and some applications}

Let $G$ denote a lattice or, more generally, an undirected connected $\kappa$-regular graph (i.e. a graph, every of whose vertices is linked to $\kappa$ neighbours). Let $v\in G$ denote the vertices or sites and the links between sites $(i,j)$ within $G$ be described by an adjacency matrix $\{m_{i j}\}_{1\leq i,j \leq |G| }$. For simplicity $m_{i j}\in \{0,1\}$ and $m_{i i}=0$. In the following $G$ will also be called base graph.  A vertex $v\in G$ can either be occupied or empty, i.e., have occupancy $s(v)=1$ or $s(v)=0$, respectively. Empty sites will be called holes. A vertex $v$ with label $i$ has degree $\partial v=\sum^{|G|}_{j=1} m_{i j}\leq \kappa$. We propose the following algorithm for  an evolution of occupied and empty sites on $G$:
\begin{itemize}
\item [(i)]  Throw in $I$ occupied vertices, i.e. select $i$ holes randomly and let them become occupied. This supply of $I$ new occupied vertices will be called influx.
\item [(ii)] Check the neighbourhood of every site $v\in G$. If $v$ is occupied and has degree $\partial v$ greater than $t_{\text{u}}$ or less than $t_{\text{l}}$ the vertex $v$ will be set empty in the next timestep. That is: $s_t(v)=1\to s_{t+1}(v)=0$ if $\partial v <t_{\text{l}}$ or $\partial v > t_{\text{u}}$. The update of (i) and (ii) will be parallel.
\item [(iii)] Iterate (i) and (ii). 
\end{itemize}
The threshold values $t_{\text{l}}$ and $t_{\text{u}}$ are model parameters, $t_{\text{l}}=0$ corresponds to no lower threshold at all. Thus, the main characteristics of the algorithm is a window of allowed degrees. Hence we call it window algorithm.
The graph of all occupied vertices after iteration $t$ is finished will be denoted by $\Gamma_t\in G$ and $n_t=|\Gamma_t|$ be called its population.

Note, that in the first instance unlimited growth of the network is prevented by the upper threshold $t_{\text{u}}$. Once a stable structure has been established, there are no vertices violating the minimum degree rule unless vertices are taken out because they got too many neighbours by the last influx. On the other hand, while $t_{\text{u}}$ leads to an instantaneous removal of vertices due to the most recent influx, the lower threshold $t_{\text{l}}$ can cause avalanches and is thus responsible for a memory of a perturbation which may last over several iterations. Such avalanches are thought to adjust the network structure.

These considerations make it also plain that for $t_{\text{u}}=t_{\text{l}}$ and $t_{\text{u}}=t_{\text{l}}+1$ no long lasting populations can arise. In the first case, only a $t_{\text{l}}-$ regular graph $\Gamma_t$ could be formed. Then already a small local reorganization would lead to an avalanche extinguishing the whole graph. In the second case, every vertex of $\Gamma_t$ is critical in a sense that it is prone to become removed by a small perturbation caused by the influx.

Our main motivation to study such a kind of dynamics for the evolution of networks comes from trying to model idiotypic networks (INWs). The problem, however, appears to be far more general. Two other imaginable applications will also be sketched briefly: evolving networks of interacting species and, for conceiving  a vivid picture of the above algorithm, a toy system of organzing coins in a box.

\subsection {Idiotypic networks}
\label{modelinginws}
Immune response against a broad class of antigen (bacteria, viruses, ...) is triggered by antibodies and B-cells (which carry only one type of antibody on their surface). Schematically this can be sketched as follows. Perchance the antigen encounters a shape- and charge-complementary antibody. Antibody and antigen thus form a complex which marks the antigen as hostile so that it will be removed by another functional group of cells (e.g. killer cells). If B-cells, which carry exactly this type of antibody on their surface come into contact with such an antigen they become stimulated, multiply into a clone and finally become `production units' for this specific type of antibody.

It is the basic idea of idiotypic networks \cite{Jerne} that antibodies can not only recognize antigen, but also complementary anti-antibodies (the specificities of both are then also called idiotypes). This leads to stimulation of the respective B-cells and thus causes a kind of a permanent self-balanced immune activity independent of hostile antigen.

These mutually recognizing idiotypes build the so-called idiotypic network. Its vertices or nodes are represented by idiotypes, its links by functional interactions between them. INW's are thought to play a role in, e.g., the preservation of idiotypic memory \cite{Jerne}, the prevention of autoimmunity \cite{Varela1,Behn1}, self-non-self discriminization and cancer control.

The interaction of idiotypes within the network can be described by Lotka-Volterra like dynamics, i.e.
\begin{align}
\label {LV}
\frac{dx_i}{dt}=x_i\left(-\gamma_i + f\left(\sum_j m_{i j} x_j\right)\right)+\xi_i,
\end{align}
where $i$ is an index labelling different idiotype populations, $x_i$ the concentration of idiotype $i$, $\gamma_i$ its inverse lifetime, $f$ a so-called proliferation function describing the stimulaton and death of B-cells and antibody production by stimulated B-cells, and $\xi_i$ an influx rate of new idiotype species from the bone marrow. Since secreted antibodies and those on the surface of B-cells are not explicitly distinguished a description of INW dynamics as by Eq. (\ref{LV}) belongs to the category of A-models \cite{Ueb}. Bone marrow production is thought to be uniformly distributed and random, i.e., there is a uniform probability for every idiotype that a member of its population is produced per unit of time.

It is a common approach to describe an idiotype or vertex $v$ of the INW by a bitstring $(v_1,...,v_d)$, $v_i\in \{0,1 \}$ of length $d$ \cite{Farmer}. Reasonable estimates for the number of possible idiotypes yield a bit-chain length of the order $d\approx 35$ \cite{Ueb}.  Idiotypes interact, if they are complementary. This is modelled by introducing `matching-rules' which define when bit-strings are connected in the above sense of complementarity. For instance, idiotypes can be said to interact if they are nearly exactly complementary or --in other words-- the respective bit-strings match allowing for one deviating position, a mismatch, only. Generally, allowing bit-strings with $m$ mismatches to interact, one obtains a $\kappa = \sum^m_{k=0} \begin{pmatrix} d \\ k\end{pmatrix}$-regular graph $G^{(m)}_d$ whose links are described by an adjacency matrix with elements:
\begin{align}
\label{mismatcheq}
m^{(m)}_{i j}= \begin {cases} 1 & \text{if } d_{\text{H}}(i,\overline{j}) \leq m \\ 0 & \text{otherwise} \end {cases}.
\end{align}
Reaction affinities are the stronger the better the matching of the corresponding idiotypes is. In this respect, in the sequence $G^{(1)}_d\subset G^{(2)}_d\subset ...\subset G^{(m)}_d$ gradually smaller affinities are taken into account.

Sometimes it is also convenient to employ another view at the link structure of $G^{(m)}_d$. In the frame of these bitstring models every link is associated with a bit-operation $L$, which, applied to one of the vertices that the link connects, yields the vertex at its opposite end. For links between vertices with ideally complementary bit-chains $i$,$j$ we have $j=L_0(i)$ ($i$ is the inverse of $j$), for `one-mismatch-links' $j=L_{k_1}(i)$ ($i$ is obtained by inverting $j$ and changing the bit at position $k_1$), for `two-mismatch-links' $j=L_{k_1 k_2}(i)$, etc. Thus, in this sense links are also labelled.

Using a discrete version of Eq. (\ref{LV}) idiotypes are either present ($x_i=1$) or absent ($x_i=0$). An idiotype $v$ can survive, if it on the one hand receives at least a minimum amount of stimulation from the network, but is not overstimulated on the other hand. Too high concentrations of antibody lead to each receptor of the B-cell being bound to only one antibody. Since cross-linking of several antibodies is required in order to stimulate a B-cell \cite{Ueb} this leads to suppression of this idiotypes population. A more careful analysis yields that it is reasonable to assume that the proliferation function $f$ is log-bell shaped. Then, from the steady state conditions of Eq. (\ref{LV}) one obtains a rough idea of the maximum and the minimum degree, $t_{\text{l}}$ and $t_{\text{u}}$. Normally, one matching antibody specifity is sufficient to cope with an antigen. Therefore in this model $t_{\text{l}}=1$ which defines the timescales for each iteration step.

\subsection {Interacting species}
A widely used approach to describe the interactions of species (macroscopic organisms, cells, reacting molecules, ...) is by Lotka-Volterra like dynamics of the type set forth in Eq. (\ref{LV}) \cite{Svireshev}. The index $i$ now distinguishes different species, $x_i$ denotes their concentrations, $-\gamma_i > 0$ the net effect of death and birth rate, $m_{i j}$ the adjacency matrix describing their interaction structure, and $f$ a function modeling  resource competition and the effect of predator-prey relations. Now, a species does not require a minimum amount of stimulation but is still going to be restrained by the effects of resource competition and prey. Thus, $t_{\text{l}}=0$ and $t_{\text{u}}>0$. $\xi$ in this context models a kind of mutation rate, which describes the appearance of new species. Unlike in the case of the bone marrow production, $\xi$ is not uniform, but restricted to the neighbourhood of already existing species and also much smaller.
Though probably hard to make out, $G$ and $m_{i j}$ describe the graph of all conceivable species and their potential interactions.

\subsection {Coins in a box}
Imagine a box and a heap of coins put inside initially. Then the box will be shaken, all overlapping coins be taken out and randomly generated new heaps of coins be placed on randomly selected empty areas of the box. This procedure will be iterated several times. Clearly, in the above algorithm, the box corresponds to the base graph $G$, coins correspond to occupied vertices on $G$, and empty areas to holes. Overlapping coins have too many neighbours, thus too high a degree and will consequently be taken out. Since also isolated coins are allowed to remain, the situations is that of $t_{\text{l}}=0$. $t_{\text{u}}$ represents the maximum amount of coins which can be adjacent on a plane without overlap.\\

\section {Analyzing the dynamics}
\label{analyzingthedynamics}

\begin{figure}
 \begin{center}
  \hspace{0.1cm}
  \epsfxsize 7 cm
  \epsffile {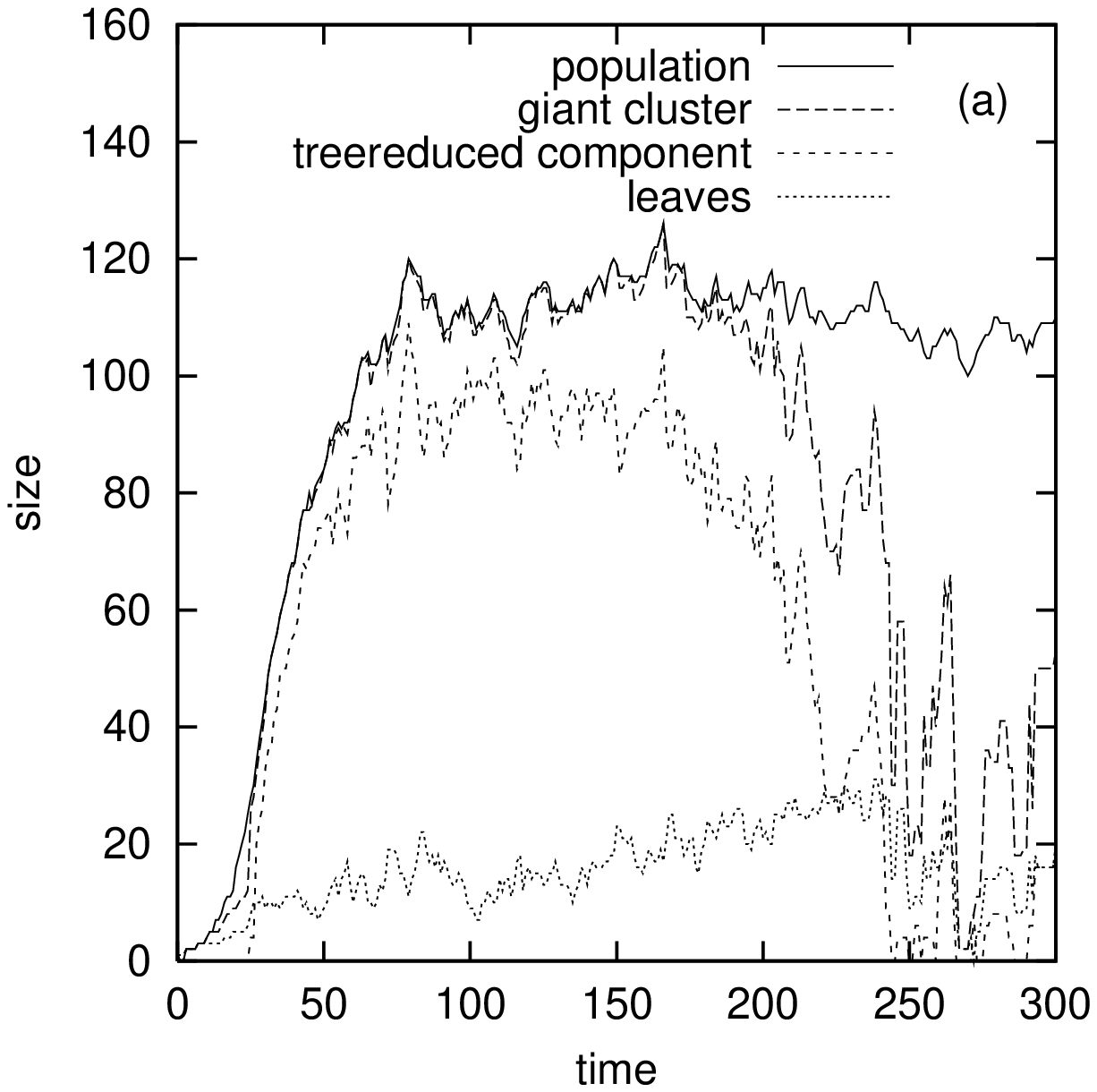}\\
  \hspace{0.1cm}
  \epsfxsize 7 cm
  \epsffile{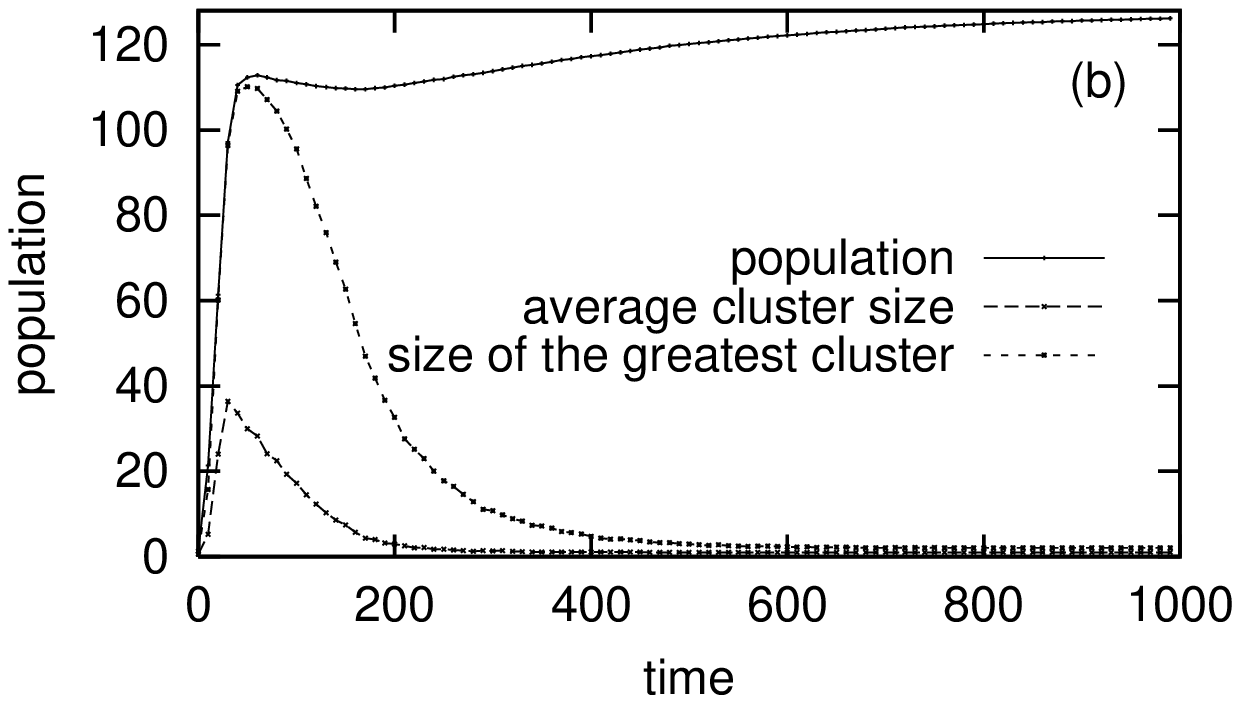}
\caption {(a) A sample trajectory for the population $n_t$, the giant cluster, the 2-tree-reduced component of the giant cluster (`backbone') and its 2-leaves, see text. Parameters are $d=12$, ($t_{\text{l}},t_{\text{u}}$)=($1,7$) and $I=10$. (b)  Averaged trajectories of the population, the size of the greatest cluster and the average cluster size vs. time. Averages have been taken over 1000 independent runs, parameters are $d=8$, ($t_{\text{l}},t_{\text{u}}$)=($1,5$), $I=6$.}
\label {FIG1}
 \end{center}
 \end{figure}

In this section we focus on analyzing the population dynamics and understanding corresponding changes in the network structure. Figure \ref{FIG1}a shows a sample trajectory for the initial 300 timesteps of the population dynamics for ($t_{\text{l}},t_{\text{u}}$)$=$($1,7$) and $I=10$ on a base graph $G^{(1)}_{12}$. As there seems to be a generic evolution with time it makes sense to look at averaged trajectories, i.e. $\langle n_t\rangle=\sum_n np_t(n)$ where $p_t(n)$ denotes the probability that after $t$ iterations $n$ occupied vertices will be present (see Fig. \ref{FIG1}b).

Clearly, several phases of the dynamics can be distinguished. First, it takes a certain time till accidentally at least a connected pair of occupied vertices is formed which then functions as a germ for further network growth.

To elucidate the structure of this germ for $t_{\text{l}}>1$ we define three subsets of a cluster $C$: (i) its (generalized) $t_{\text{l}}$-leaves $l_{t_{\text{l}}}(C)$, (ii) its (generalized) $t_{\text{l}}$-stem $s_{t_{\text{l}}}(C)$ and (iii) its (generalized) $t_{\text{l}}$-tree-reduced component $r_{t_{\text{l}}}(C)$. Leaves of the first generation are vertices of $C$ with less than $t_{\text{l}}$ neighbours, i.e. $l^{(1)}_{t_{\text{l}}}(C)=\{v\in C|\partial v<t_{\text{l}} \}$. Then, leaves of the second generation $l^{(2)}_{t_{\text{l}}}(C)$ are vertices of $C\backslash l^{(1)}_{\text{l}}(C)$ with less than $t_{\text{l}}$ neighbours. Analogously for $n=2,3,...$, $l^{(n)}_{t_{\text{l}}}=\{v\in C\backslash l^{(n-1)}_{t_{\text{l}}} |\partial_{|C\backslash l^{(n-1)}_{t_{\text{l}}}}v<t_{\text{l}}\}$, where $\partial_{|C\backslash l^{(n-1)}_{t_{\text{l}}}}v$ denotes the number of occupied neighbours of $v$ in $C\backslash l^{(n-1)}_{t_{\text{l}}}$. Thus, the tree-like component of $C$ is $t_{t_{\text{l}}}(C)=\bigcup^{\infty}_{j=1}l^{(j)}_{t_{\text{l}}}(C)$, the stem $s_{t_{\text{l}}}=t_{t_{\text{l}}}-l^{(1)}_{t_{\text{l}}}$ and the tree-reduced component $r_{t_{\text{l}}}=C\backslash t_{t_{\text{l}}}$. For $t_{\text{l}}=2$ these notions coincide with the usual definitions in graph theory.

Consider the application of the window-algorithm with lower threshold $t_{\text{l}}$. If not sustained by random influx, the $t_{\text{l}}$-stem of a cluster is prone to successively fall victim to cascades in which the $t_{\text{l}}$-leaves are taken out first. Vertices which are critical, i.e. are still allowed, but will be removed if they lose a neighbour represent the $(t_{\text{l}}+1)$-leaves. Contrariwise, the $(t_{\text{l}}+1)$-tree-reduced component is the subset not endangered by avalanches caused by critical vertices.

Apparently, the tree-reduced component of a cluster is invariant under step (ii) of the window-algorithm. So, for $t_{\text{l}}>1$ in the above context the germ which for $t_{\text{l}}=1$ is just a connected pair of occupied vertices corresponds to a cluster with a nontrivial $t_{\text{l}}$-tree-reduced component. In case of, e.g.,  $t_{\text{l}}=2$ this is a 4-loop, for $t_{\text{l}}=3$ a cube etc. The higher $t_{\text{l}}$ the more `organized' a germ is demanded and the longer it takes till it is formed perchance.

Initially, as this germ is still small it grows only slowly, since some of the new vertices are short of neighbours and thus are immediately taken out again. Then as a second phase of the dynamics a phase of almost linear growth of the population is entered.  During this period almost all freshly thrown in vertices survive, since the network is already almost dense (thus providing enough neighbours) but still most of the vertices have only small degree. As the number of vertices with higher degrees increases the growth of the population then abates and reaches a maximum. It can also be seen from Fig. \ref{FIG1}b that during this phase almost all occupied vertices belong to one giant cluster. At the same time $(t\approx 150)$ as the maximum is reached, the giant cluster decays and small components start to split off. Interestingly, whereas in the initial stages of the dynamics the number of critical vertices is always of the same size as the influx, preceding the breakdown of the giant component more and more critical vertices are assembled (cf. Fig. \ref{FIG1}a). The maximum number of critical vertices is assumed at the same time ($t\approx 290$) as the giant cluster breaks down.

The third phase now ensuing is marked by the complete decay of the giant cluster and a coinciding drop of the population. As soon as this is achieved the smaller fragments start becoming rearranged, again associated with an increase in the population.

Finally, the system saturates into a stationary state around which fluctuations occur. The higher $t_{\text{u}}$ the more the population is allowed to grow initially and the more expressed the above described behaviour becomes. Independent of the upper threshold always half of the vertices are occupied in the final state.
Recurring to clustering properties it turns out that average as well as maximum cluster size tend to $2.0$. Thus, since there can't be a cluster consisting of more than two occupied vertices this proves that for long times $\Gamma_t$ completely decays into 2-clusters.

Holes play an important role in the systems approaching the steady state. A hole $h$ will be called stable if it has degree $\partial h > t_{\text{u}}$. Clearly, unless it loses its property to be stable, a stable hole cannot become occupied. Reverting a stable hole back into a `normal' hole demands the removal of $\partial h-t_{\text{u}}$ of its occupied neighbours, which itself needs a rearrangement in the 1-neighbourhood of these vertices. Thus, two characteristics of stable holes become obvious: (i) a hole is the more stable the higher its degree and (ii) whereas the change of a normal hole into an occupied vertex demands only a rearrangement at exactly the holes place, the occupation of a stable hole requires a rearrangement of its whole 2-neighbourhood and needs at least two iterations to be completed.

\begin{figure}
 \begin{center}
  \epsfxsize 7 cm
  \epsffile{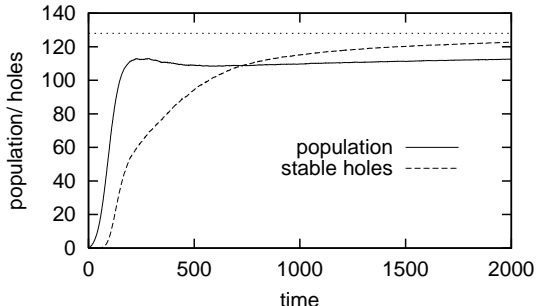}
\caption {Averaged trajectory (over 1000 independent runs) of the number of stable holes and of the population for $d=8$, ($t_{\text{l}},t_{\text{u}}$)$=$($1,5$), $I=2$. The dotted line indicates half the system size $|G^{(1)}_8|/2=128$.}
\label {FIG2}
 \end{center}
 \end{figure}

Figure \ref{FIG2} shows an averaged trajectory for the evolution of the number of stable holes and the population size. Surprisingly, even during the phase of declining population (which means that fewer occupied vertices to make a hole stable are available) the number of stable holes increases till finally in the steady state all holes tend to be stable. This can be understood from the above considerations. As described above in detail, the reversion of a stable hole into a normal hole requires occupied vertices in the neighbourhood of the hole to be removed. However, the removal of vertices becomes more difficult the more stable holes are in their neighbourhood. Thus, apparently, stable holes situated in each others (up to 2-) adjacency can promote mutual stability. As a consequence of this a formation of almost `frozen' domains which are made of mutually stabilizing stable holes can be expected. These domains compete with each other and finally one of them prevails and fills the complete system. Indeed, the arrangement of occupied vertices and holes in the steady state exhibits patterns which will be discussed more closely in the next section.

\section {The stationary state}
\label{thestationarystate}
In order to characterize the network structure of the stationary state every vertex $v\in G$ is assigned a mean occupancy rate $\bar{s}(v)=1/(T_1-T_0)\sum^{T_1}_{t=T_0} s_t(v)$ and a switch rate $r(v)=1/(T_1-T_0)\sum^{T_1}_{t=T_0} (1-\delta_{s_t(v), s_{t+1}(v)})$. Mean occupancy rate and switch rate give a measure of how many times a vertex is occupied or changes its occupation during the time interval $[T_0, T_1]$. $T_0$ will be chosen such that the system has already reached the stationary state. This method provides a way of obtaining an average picture of the occupation of $G$.
\begin{figure}
 \begin{center}
  \epsfxsize 7 cm
   \epsffile{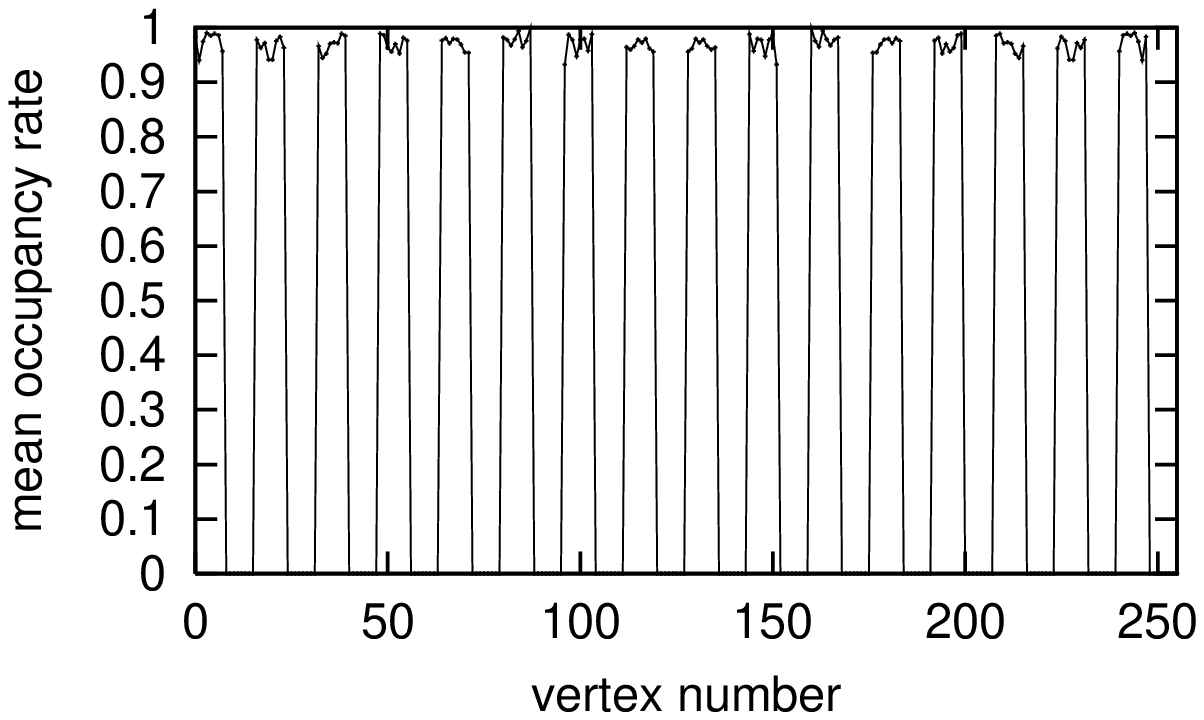}
   \epsfxsize 7 cm
   \epsffile{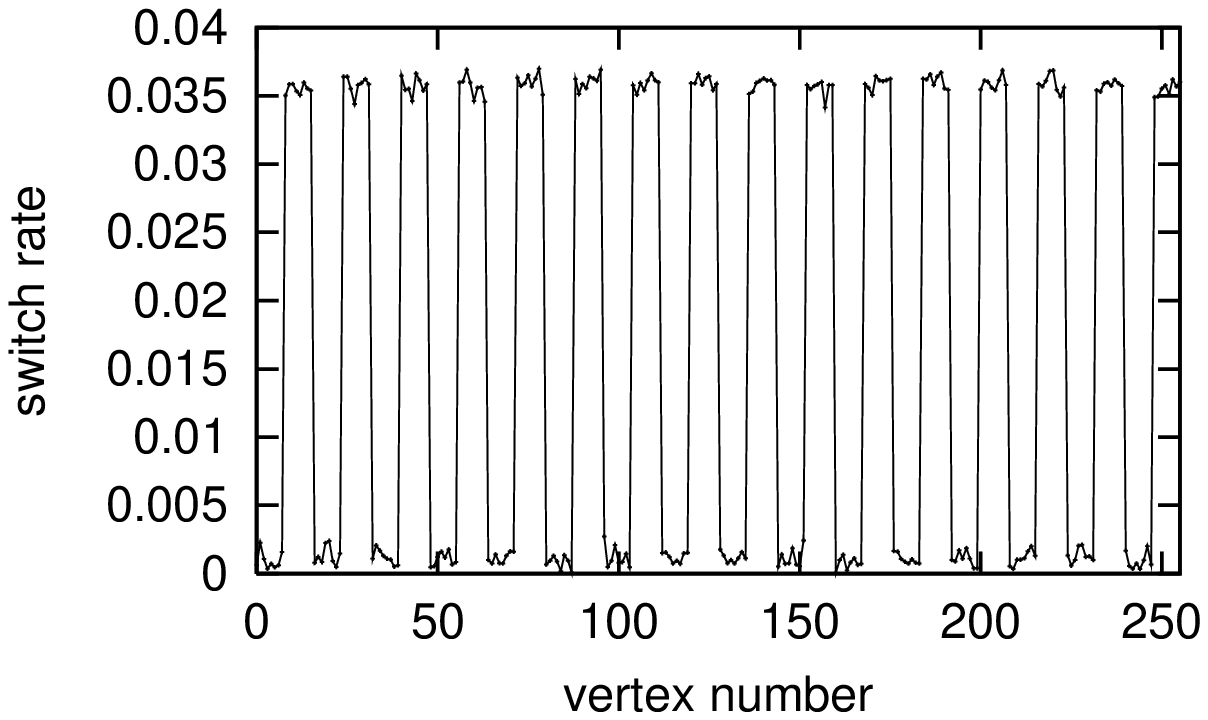}
   \vspace* {0.5cm}
\caption {Mean occupancy rate and switch rate for every vertex after a simulation  on $G^{(1)}_8$ with parameters ($t_{\text{l}},t_{\text{u}}$)$=$($1,4$), $I=5$. The relaxation time was chosen $T_0=10^4$, then averages over $T_1=9\times10^4$ timesteps have been taken. Bit-strings $v=(v_1,...,v_8)$ are uniquely represented by integers $z=\sum^7_{i=0}v_i2^{i}$.}
 \label {FIG3}
 \end{center}
 \end{figure}
Figure \ref {FIG3} shows the mean occupancy rate and switch rate of every vertex after a simulation on $G^{(1)}_8$. Clearly, vertices with high mean occupancy rate have a low switch rate and vice versa. Thus, vertices which are frequently occupied tend to be occupied almost permanently; seldom occupied ones to be almost always holes.

To elucidate the structure of the network formed by the highly active vertices we define the subset $S(a)=\{v\in G|\bar{s}(v)<a \}$ containing all vertices which are more frequently occupied than with rate $a$.

\begin{figure}
 \begin{center}
  \epsfxsize 7 cm
   \epsffile{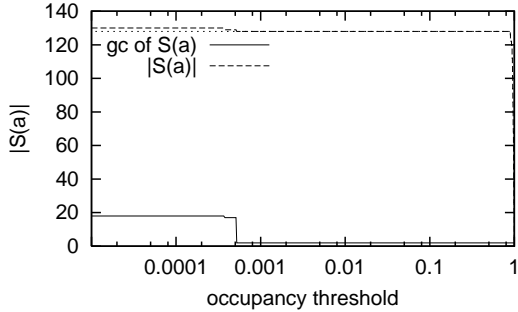}
   \vspace{0.5cm}
\caption {Number of vertices belonging to $S(a)$ (see text) and the size of the greatest connected component of $S(a)$ as a function of the threshold $a$. Data taken from simulations on $G^{(1)}_8$ with the window ($t_{\text{l}},t_{\text{u}}$)$=$($1,4$) and influx $I=5$. Due to the logarithmic scale the first point showing a drop of $|S(a)|$ from $|G|=256$ to 130 is left out. The dotted line indicates the level $|S(a)|=128$. Data have been assembled for $T_1=9\times 10^4$ iterations after a relaxation time $T_0=10^4$.}
 \label {FIG4}
 \end{center}
 \end{figure}

In Fig. \ref{FIG4} the number of vertices belonging to $S(a)$ and the greatest connected cluster of $S(a)$ are shown as a function of the threshold value $a$. Due to the logarithmic scale a first drop from $|S(a)|=256$ to $|S(a)|=130$ at $a=0$ is omitted. Two further thresholds become apparent: one at $a_1\approx 3\times 10^{-4}$ where the size of the greatest component of $S(a)$ rapidly drops to size 2 and a second one at $a_2 \approx 0.893$ which marks the decline in the distinction of high- and low mean occupancy vertices seen in Fig. \ref{FIG3}.
\begin{figure}
 \begin{center}
  \hspace {0.1cm}
  \epsfxsize 4 cm
   \epsffile{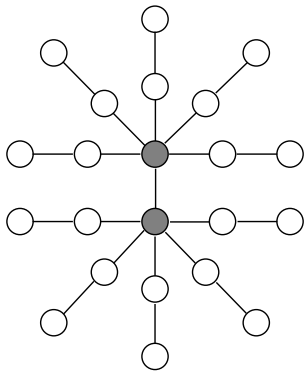}
   \vspace{0.5cm}
\caption {Illustration of the cluster structure of the defect caused by the two vertices with mean occupancy rate $\bar{s}\approx \bar{s}_1$ (drawn in grey) for $\kappa=6$. These vertices are in the centre of a star-like cluster of other low-mean occupancy vertices from $L_1$ (empty circles). For $a>a_1$ they no longer belong to $S(a)$ and thus their removal causes a decay of this cluster into ten 2-clusters of holes.}
\label {FIG4a}
 \end{center}
 \end{figure}
In order to get rid of fluctuations occuring during the iterations, one can distinguish a set of high mean occupancy vertices $S_H=S(\bar{s}_1)$ (which can be considered as permanently occupied) and a set $S_L=G\backslash S_H$ (corresponding to permanent holes) of vertices with low mean occupancy. From Fig. \ref{FIG4} one derives further that $S_L$ decays also into two groups, namely, a set $S_{L_1}$ of 128 vertices with mean occupancy 0 and two vertices (seen as a drop of $|S(a)|$ from $130$ to $128$ at $a=a_1$) with somewhat larger mean occupancy $a$ slightly smaller than $a_1$. These 2 connected vertices having each $\kappa-1=8$ neighbours from $S_{L_1}$ are in the centre of a star-like cluster of low mean occupancy vertices (cf. Fig. \ref{FIG4a}). They represent a vertex-pair normally belonging to $S_H$ which have accidentally been taken out and not got refilled yet. Thus they form a defect in the expression of a pattern structure (see section \ref{defectssection}).

A further clustering analysis of $S_H$ and $S_L$ reveals that both subsets of $G$ consist only of 2-clusters. Therefore, every vertex of $v\in S_L$ has exactly $\partial v=\kappa-1$ neighbours from $S_H$ and thus forms a stable hole of maximum possible stability. This justifies the above notion of a frozen domain.

In case of $t_{\text{u}}=\kappa-1$ a holes being stable would require at least $\kappa=t_{\text{u}}+1$ of its neighbours to be occupied. Due to the 2-cluster nature of the pattern, however, only $\kappa-1$ neighbours are available, so that holes cannot be stable in this case. Patterns then become extremely transient in the sense discussed below in section \ref{transitionregion}. 

\subsection {Metastable patterns and pattern stability}
\label{patternstability}
The actual steady state pattern, classified by groupings of vertices with clearly distinguished mean occupancy rates, depends on the influx $I$. If $I$ is very small compared to the system size $|G|$ also patterns different from the 2-clustered pattern appear. The variety of such patterns is very abundant for $t_{\text{l}}=0$, becomes smaller the more restrictive the window is and increases with decreasing ratio $I/|G|$. 

They prove to be metastable, i.e. relax to a more stable pattern and finally to the 2-cluster pattern on very long timescales. Interestingly, for $t_{\text{l}}=0$ both sets $S_H$ and $S_L$ prove to have exactly alike properties and thus, the system is symmetric against an exchange of occupied vertices and holes \cite{FN0}.

\begin{figure}
 \begin{center}         
  \hspace{0.1cm}
\epsffile {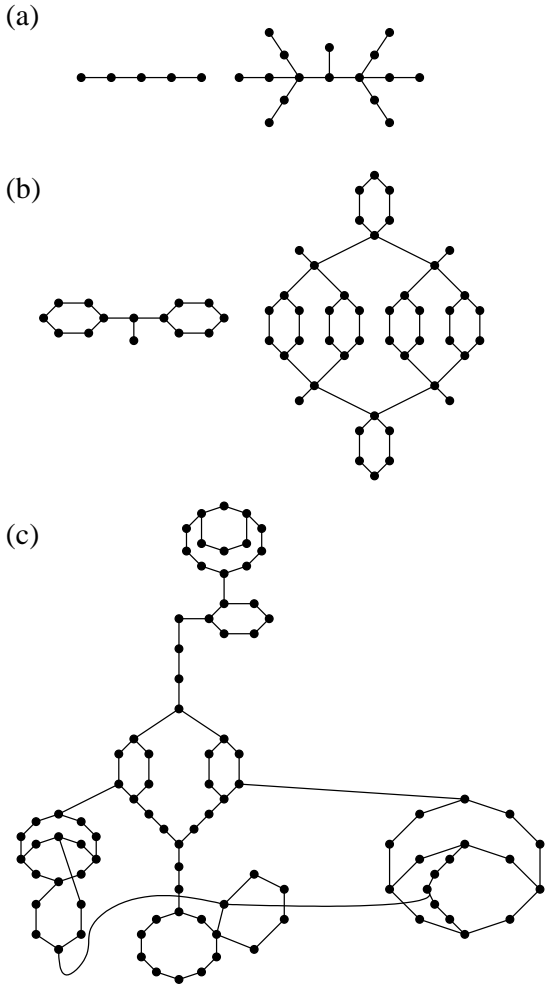}
\end{center}
 \vspace {0.5cm}
 \caption {Nontrivial clusters  which appear in metastable patterns for $I=5$, ($t_{\text{l}},t_{\text{u}})$=($0,7$) on $G^{(1)}_{11}$. (a) 5- and 16-clusters from a pattern built by 192 singletons, 64 5- and 32 16-clusters. (b) 14- and 44-clusters from a pattern with with 208 1-, 48 5-, 16 14- and 8 44-clusters. (c) 2-Tree-reduced component of the 203-cluster from a pattern consisting of 152 1-, 12 2-, 12 3- and 4 203-clusters (2- and 3-clusters form chains). In the cases (a) and (b) larger clusters contain structures which resemble the smaller clusters. Larger loops, especially 6-loops and trees play a distinguished role. In all cases the dimension of the cluster structures is remarkably low.}
\label {FIG5}
 \end{figure}

Figure \ref{FIG5} gives examples for such a metastable pattern together with a visualization of the ismomorphic graph structures of $S_H$ and $S_L$. Typically, they consist of a large number of small clusters and few larger components. In the organization of the network structure of subclusters of $S_H$ and $S_L$ basic units (e.g. 6-loops in Fig. \ref{FIG5}) play a role. Nevertheless the majority of the clusters turns out to be asymmetric.

In this context a comparison to {\it random} graphs on $G$ illustrates the very high organization of these patterns. A random graph $\Gamma \subset G$ (constructed by occupying vertices of $G$ with a uniform occupation probability $p$) will contain almost always only one giant connected component for $p>p_c\sim 1/\kappa$ \cite{BBM}. Here, multiple giant components exist, although the fill ratio $|H|/|G|=0.5$ far exceeds the percolation threshold.

Stability of a pattern is determined by the stability of its stable holes. Thus, a (rough) measure for the stability of a pattern is, e.g., the average degree of its holes $\langle \partial v \rangle_{v\in S_L}$. It turns out that all these metastable patterns have an average hole degree slightly smaller than that of the 2-clustered configurations.

\begin{figure}
 \begin{center}
  \epsfxsize 7 cm
   \epsffile{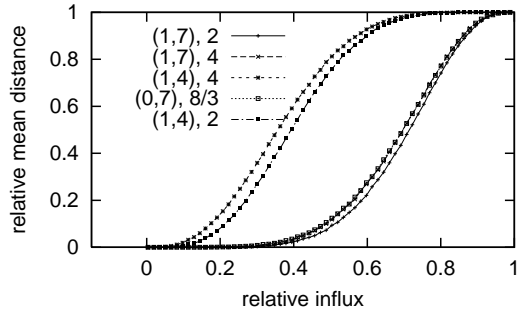}
   \vspace{0.5cm}
\caption {Relative mean distance $\langle d(P,P')\rangle/|L|$ as a function of the relative influx $\tilde{I}/|L|$ for patterns which arise on $G^{(1)}_9$. Different lines correspond to patterns with different mean cluster sizes $\langle c\rangle$ which emerge for the given combinations of thresholds $(t_{\text{l}},t_{\text{u}})$, see text. For given relative influx patterns are the more stable the smaller their relative mean distance.}
 \label {FIG6}
 \end{center}
 \end{figure}

Figure \ref{FIG6} shows another way to quantify stability. For this purpose we apply a test influx $\tilde{I}$. More specifically, a pattern $P$ of holes and occupied vertices is taken, a number $\tilde{I}\leq |L|$ of its holes $h$ belonging to the low mean occupancy set $S_L$ randomly selected and occupied, the window algorithm applied and the distance $d(P,P')=\sum_{v\in G}|s_P(v)-s_{P'}(v)|$ between the original pattern and the resultant graph $P'$ measured. Intuitively, $\tilde{I}$ is a measure for a perturbation which is applied to the pattern. A small mean deviation $\langle d(P,P')\rangle$ means a high stability against random influx $\tilde{I}$. For growing system sizes the sigmoid function $\langle d(P,P')\rangle/|L|$ tends to a step function.

Confirming the above ideas, Fig. \ref{FIG6} visualizes two tendencies: (i) an upper threshold $t_{\text{u}}<\kappa-1$ closer to $\kappa-1$ \cite{FN1} and (ii) a higher mean degree of holes $\langle \partial h\rangle_{h\in S_L}$ promote the stability of patterns. Indeed, patterns with the same mean hole degree have almost the same stability characteristics $\langle d(P,P')\rangle/|L|$. For example, the patterns with $(t_{\text{l}},t_{\text{u}})=(0,7), \langle c\rangle=8/3$ and $(1,7), \langle c\rangle=4$ whose curves coincide in Fig. \ref{FIG6} have both a mean hole degree $\langle \partial h\rangle_{h\in S_H}=8.5$.

Due to the high supply of new idiotypes from the bone marrow these patterns are without interest for the purpose of modelling immune networks. However, it could be conceived that they prove interesting in the context of evolving species interaction networks where the influx as the mean number of new species per time unit is small in comparison to the overall network size.

\subsection {Stable base configurations}
\label{statedescription}
The 2-clustered patterns are distinguished in two respects: (i) they are the only patterns evolving if the influx is high and (ii) they are stable, i.e., if once such a pattern is assumed the system will not relapse to one of the metastable patterns. Therefore we call it stable base configurations.

The complete decay of both hole- and vertex-configurations into 2-clusters allows it to calculate the multiplicity of these base configurations. It turns out that the global structure of a base configuration is already determined locally by fixing the occupancy $s_0$ of one connected pair of vertices. By the knowledge of the 2-cluster structure of $S_H$ and $S_L$ one then knows that the vertices in the neighbourhood of this pair have occupancy $s_1=1-s_0$. Thus, also the occupancy of their 2-cluster mates is fixed, which in turn determine the occupancy of vertices in their neighbourhood and so on. By this argument every vertex' occupancy is uniquely determined since $G$ is connected. All base configurations constructed in this way have been found to occur and are isomorphic to each other in the sense, that without labelling vertices they can't be distinguished.

As a consequence of the above argument, fixing $s_0$ and chosing one arbitrary vertex $o\in G$ and a link associated with a bit-operation $P$ leading to a two-cluster mate of $o$, $P(o)$, base configurations can be denoted by $B_{(s_0, P)}$. Hence, the number of possible base configurations is $2\kappa$. The system is degenerate, though as $2\kappa/|G|\to 0$ for $|G|\to \infty$ not macroscopically.

From a biological point of view it appears appealing to identify 2-clusters with idiotype-antiidiotype pairs the antiidiotype playing the role of an internal image (memory) of a previously encountered antigen \cite{Jerne,Bona,Ruiz}. The 2-clusters found here, however, are `coherent' in a fixed pattern of a stabe base configuration. This implies a maximum storage capacity $2\kappa$ which --also for large systems-- is far too few to account for experimental observations.

It appears useful to consider distances between base configurations. One finds:
\begin{align}
\label {BC}
d(B_{(s_0, P)}, B_{(s_1, Q)}) =
\begin{cases}
 |G|/2 & \text{if } P\neq Q \text { or } s_0\neq s_1 \\
 |G| & \text{if }  P=Q \text { and } s_0 = 1-s_1\\
 0 & \text {if } P=Q \text { and } s_0=s_1.
\end{cases}
\end{align}Therefore, except the `inverse configuration', base configurations differ in exactly the half of their sites from all other base configurations. For the case $P\neq Q$ or $s_0 \neq s_1$ half of the differing vertices are holes and half occupied vertices. Thus, the difference between base configuration is of the order of the system size and a change from one base configuration to another cannot take place without a major reorganization.

As the influx $I$ increases occupied vertices become more and more prone to be removed. So, it can be expected that the distinction between highly active and scarcely active vertices loses sharpness. Increasing $I$ over a threshold leads to a completely different behaviour of the system. Parameter regimes in which the system exhibits qualitatively different dynamics will be discussed in the next section.

\section{dependence on the influx $I$}
\begin{figure}
\begin{center}
  \epsfxsize 4.2 cm \hspace{0.2cm}
   \epsffile{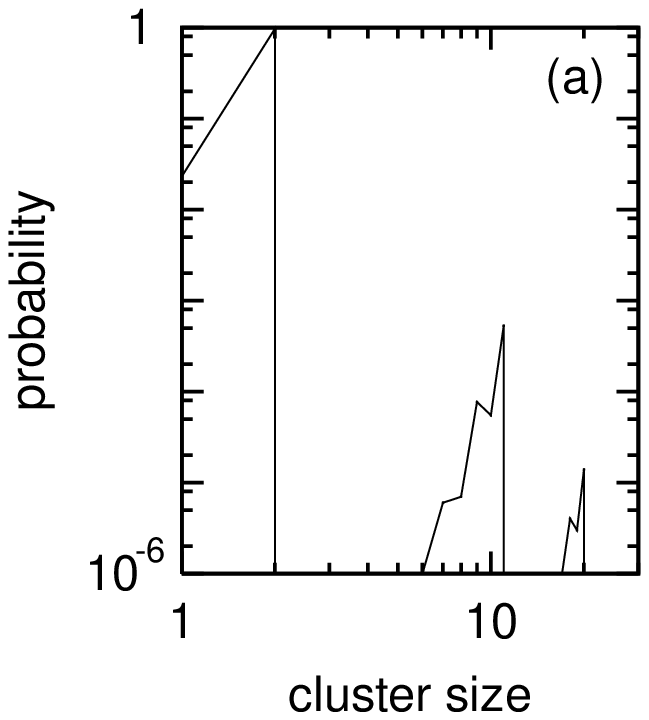} 
   \epsfxsize 4.2 cm  \epsffile{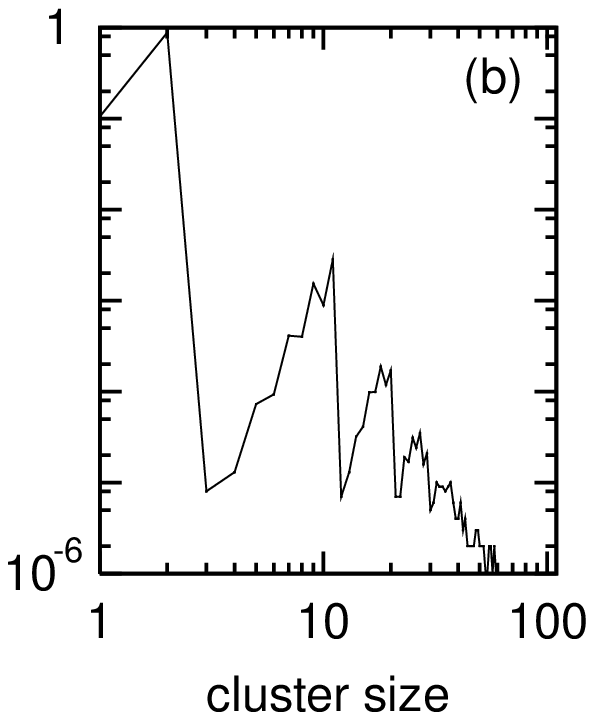} 
   \epsfxsize 4.2 cm \hspace {0.3cm} \epsffile{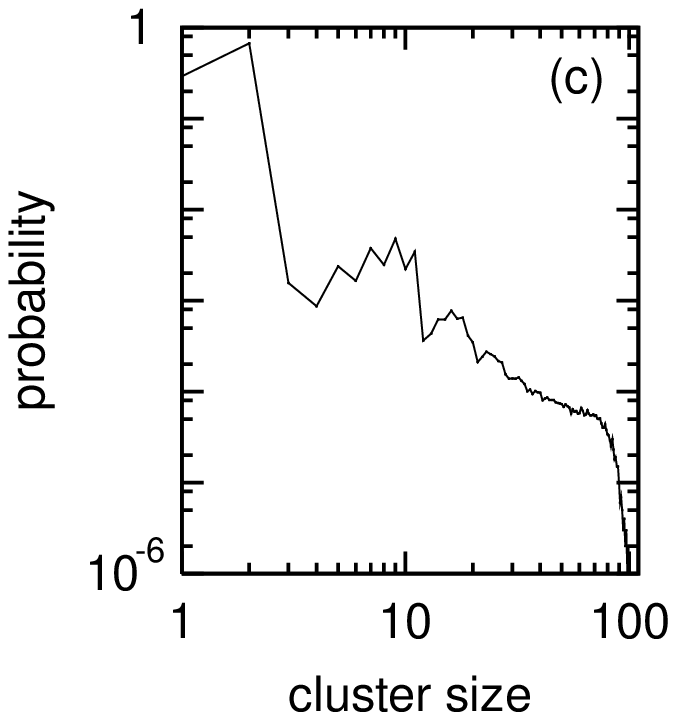} 
   \epsfxsize 4.2 cm \epsffile{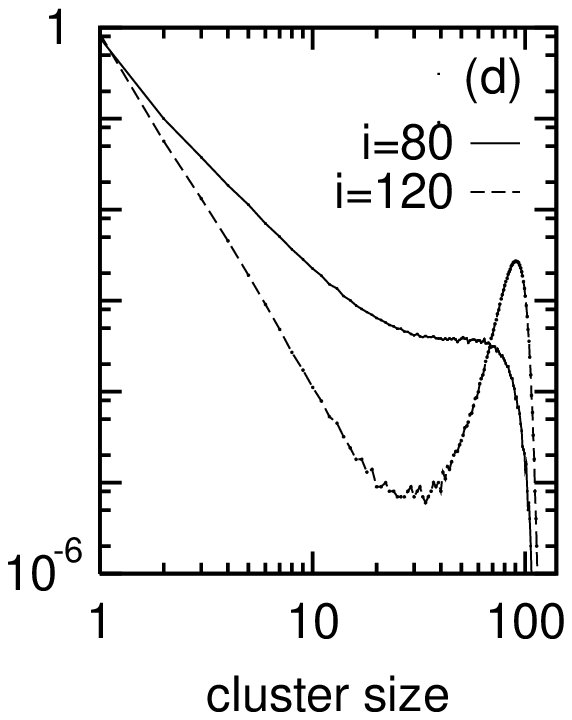}
  \vspace{0.5cm}
  \caption {Numerical data for the cluster size distribution for different influxes $I=20$ (a), $30$ (b), $40$ (c), $80$, and $120$ (d) corresponding to different regimes as described in the text. Simulations were performed on $G^{(1)}_{8}$ with ($t_{\text{l}}, t_{\text{u}}$)=($1,5$).}	
  \label {FIG9}
 \end{center}
 \end{figure}

The systems behaviour depends crucially on the clustering properties of $\Gamma_t$. To obtain a picture of the dynamics of cluster changes the time-sampled cluster size distribution over the interval $[T_1,T_0]$, $p(c)=1/(T_1-T_0)\sum^{T_1}_{t=T_0}\sum_{n_c}n_c p_t(n_c, c)$ where $p_t(n_c,c)$ denotes the probability that $\Gamma_t$ contains $n_c$ clusters of size $c$, has been investigated. Figure \ref {FIG9} shows numerical data for the cluster size distribution obtained for five different values of $I$. For small influx $I=20$ (Fig. \ref{FIG9}a) the systems behaviour is dominated by the occurence of 2-clusters. Occasionally also other clusters of sizes $c=6,...,11$ and sizes $c=16,...,20$ appear. For $I=30$ (Fig. \ref{FIG9}b) the distribution has a structured tail of series of local extrema.  Then, for $I=40$ (Fig. \ref{FIG9}c) the 2-cluster dominated structure is still expressed, but now also clusters some orders of magnitude larger are found. Finally, for $I=80$ and $I=120$ (Fig. \ref{FIG9}d), the clustering structure is dominated by (depending on $t_{\text{l}}$ and $t_{\text{u}}$) one or several giant clusters and 1-clusters (singletons) which more and more outweigh other small clusters as $I$ is further increased. $I=20$, $40$ and $80$ represent three different parameter regimes for the influx which will be treated below.
\subsection {Small $I$: A statistical approach of defects}
\label {defectssection}

\begin{figure} [b!]
\begin{center}
 \hspace{0.1cm}
\epsffile{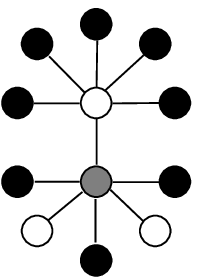} \hspace{1cm}
\epsffile{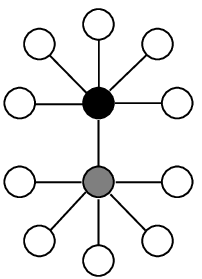}
 \vspace {0.5cm}
\caption {Illustration of `hole-defects' (left) and `vertex-defects' (right) for ($t_{\text{l}},t_{\text{u}}$)$=$($0,4$). The defective vertex is printed in grey, holes in white and occupied vertices in black. At the left hand side one of the central holes has five occupied neighbours and thus is stable. The grey hole has lost two of its neighbours due to `vertex defects'. Only three neighbours remain so that the grey hole has lost its stability and can be occupied. Both unstable and occupied holes are called hole-defects. At the right hand side the grey vertex got removed, causing thus a defect to the ideal 2-cluster pattern structure.}
\label {FIG10}
 \end{center}
 \end{figure}

For small $I$ during the initial stages of the dynamics the 2-cluster dominated pattern as described in section \ref{statedescription} is formed. Once such a pattern is established $I$ works as a perturbation which incidentally causes defects to the ideal pattern structure. These defects can be classified into two groups: `hole'-defects and `vertex'-defects (cf. Fig. \ref{FIG10}). Clearly, hole-defects are caused by $\kappa-1-t_{\text{u}}$ correlated vertex-defects and thus are much less probable than vertex defects. As long as only vertex defects occur only 1- and 2-clusters can be formed. However, once a hole defect is established, it serves as a junction for its surrounding 2-clusters and clusters of size $2\nu +1$, $\nu=1,2,...,\kappa$ are built. For example, in Fig. \ref{FIG9}a, local maxima are found for $c=7,9$, and $11$. The latter one corresponds to the most probable kind of hole defect: a formerly stable hole now is occupied and has five remaining occupied neighbours. In the case of the less probable cluster of size $c=9$ four once occupied neighbours of the hole are defective, thus four are remaining. Analogously for $c=7$. As the 2-clusters surrounding a hole defect might also be defective though less frequently also clusters of even size appear. The second series of local maxima around $c=20$ in Fig. \ref{FIG9}a can be explained by hole defects which are correlated, i.e., hole defects whose occupied 2-cluster neighbours are connected.

This microscopic picture of the process suggests a treatment borrowed from equilibrium statistical mechanics. Since defects are more likely to occur for higher influx, $I$ can be associated with a kind of a `temperature' $T(I)$. $T(I)$ will be growing monotonically with $I$. Defects are assigned an `energy' $\alpha (t_{\text{u}})$ which describes the probability that a defect is caused. Since it is the harder to remove occupied vertices the higher $t_{\text{u}}$ it is also expected that $\alpha$ grows with increasing upper thresholds $t_{\text{u}}$.

In the following we derive a statistical description for an ideal gas of defects.

\subsubsection{The case {$t_{\textrm{l}}$}=0}
The probability to cause a vertex defect depends on the number of holes surrounding the occupied vertex. In spite of this no distinction between defects of completely removed 2-clusters and singletons will be made. This is justified since for large systems an occupied vertex which remains after its 2-cluster mate has been removed is still surrounded by almost the same number of holes. Furthermore, for $t_{\text{l}}=0$ the algorithm allows the permanent existence of singletons.

The probability of having $l$ independent defects of energy $\alpha$ at temperature $1/\beta(I)$ in an ideal pattern of $N=|G|/2$ occupied vertices is
\begin{align}
\label {P1}
p(l)= \frac{1}{Z} \begin{pmatrix}N \\ l \end{pmatrix} e^{-l\beta \alpha},
\end{align}
where the partition sum $Z$ is easily determined as
\begin{align}
\label {Z}
Z=\sum^{N}_{l=0} \begin{pmatrix} N\\ l\end{pmatrix} e^{-l\beta \alpha}=(1+e^{-\beta \alpha})^{N}.
\end{align}
Then the mean number of defects is given by
\begin{align}
\label{MP1}
\langle l\rangle =-\frac{1}{\beta} \frac{\partial \ln Z}{\partial \alpha} = N-\langle n \rangle= N \frac{e^{-\beta \alpha}} {1+e^{-\beta \alpha}},
\end{align} 
$\langle n\rangle$ being the mean number of actually occupied vertices.

\begin{figure}
\begin{center}
  \epsfxsize 7 cm
  \epsffile {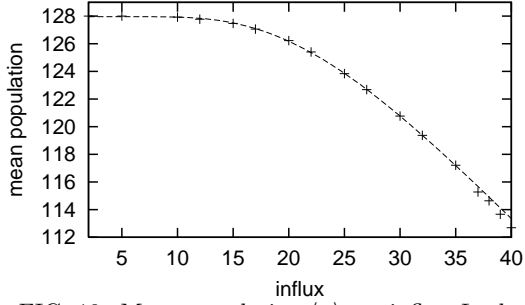}
  \caption {Mean population $\langle n\rangle$ vs. influx $I$ taken from simulations for ($t_{\text{l}},t_{\text{u}}$)$=$($0,5$) on $G^{(1)}_8$. The crosses represent simulation data, the line corresponds to a fit of $\langle n\rangle$ calculated from Eq. (\ref{MP1}). Fit parameters are $\alpha=88.14$ and $I_0=0.67$.}
 \label {FIG12}
 \end{center}
 \end{figure}

\begin{figure}
\begin{center}
  \epsfxsize 7 cm
  \epsffile {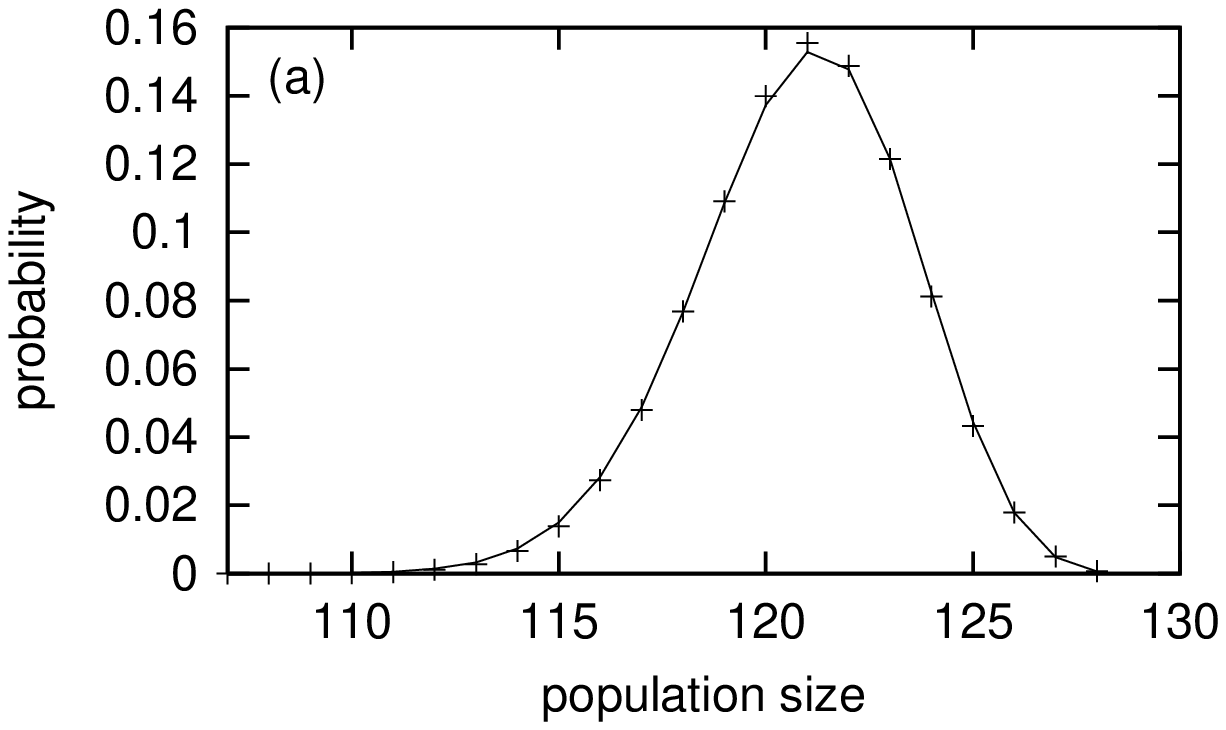}
  \epsfxsize 7 cm
  \epsffile {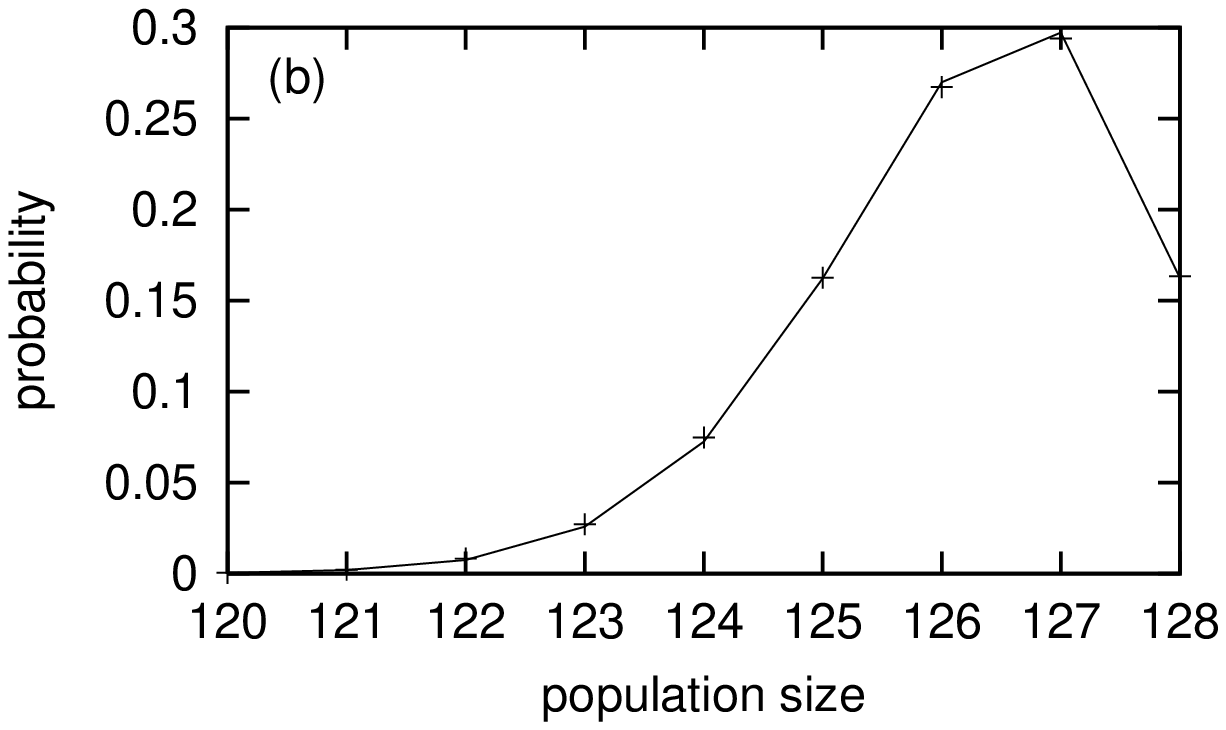}
  \caption {Probability distributions $p(n)=$$1/(T_1-T_0)$$\sum^{T_1}_{t=T_0}\sum_nnp_t(n)$ for the population size  obtained from simulations on $G^{(1)}_8$ for ($t_{\text{l}},t_{\text{u}}$)$=$($0,5$) and $I=30$ (a) and $I=20$ (b). The simulation results (crosses) are compared with $p(N-n)$ from Eq. (\ref{P1}) (lines connect calculated points) using $\alpha=88.14$ and $I_0=0.67$.}
 \label {FIG13}
 \end{center}
 \end{figure}

Comparing with data obtained from simulations it turns out that the ansatz $1/\beta=I_0+I$ holds for a broad range of different values of $I$. From a fit of the mean populations with Eq. (\ref {MP1}) one obtains $\alpha= 88.14\pm0.48$ and $I_0=0.67\pm0.18$. In the whole range up to $I<35$ mean values and probability distributions calculated by Eq. (\ref{P1}) are in good agreement with simulations (see Figs. \ref {FIG12} and \ref{FIG13}).

For very small $I$ it is by simple combinatorics possible to obtain an approximation for the `energies' fitted from the data of Fig. \ref{FIG12}. If for fixed $t_{\text{u}}$ the influx $I$ is so small that it can cause not more than exactly one vertex defect, i.e. $I<2(t_{\text{u}}+1)$, the probability per site for a defect is \cite{FN1a}
\begin{align}
\label {prob1}
p_{d}(I)=\begin{pmatrix} N \\ I\end{pmatrix}^{-1} \sum^{\kappa-1}_{j=t_{\text{u}}} \begin{pmatrix} I \\ j \end{pmatrix}\begin{pmatrix} \kappa-1 \\ j\end{pmatrix} \begin{pmatrix} N-(\kappa-1) \\ I-j\end{pmatrix}.
\end{align}
If no more than one defect can arise, the mean number of defects is given by $\langle l\rangle (I)=Np_d(I)$. Using Eq. (\ref{prob1}) one calculates $p_d(t_{\text{u}})$ and $p_d(t_{\text{u}}+1)$. A comparison with Eq. (\ref{MP1}) yields
\begin{align}
\alpha^{-1} \approx (\ln p_d(t_{\text{u}}))^{-1}- (\ln p_d(t_{\text{u}}+1))^{-1},
\end{align}
which, e.g., evaluated for $d=8$ and $t_{\text{u}}=5$ gives $\alpha=50.8$ which is of the same order as the numerical value $\alpha=88.14$. Comparing $I_0$ is not reasonable, since Eq. (\ref{MP1}) bears deficits for small influxes.


Indeed, the range of validity of Eq. (\ref{MP1}) is limited in two respects. First, although the possible number of defects for small $I$ is limited by $\lfloor I/t_{\text{u}}\rfloor$ the sum in Eq. (\ref{Z}) extends up to $N$. However, for low temperatures surplus defects are counted with only a small statistical weight. This approximation becomes worse for high upper thresholds $t_{\text{u}}$. Secondly, too many vertex defects (which then become correlated) lead to hole defects. Hole defects destabilize the 2-cluster pattern and lead to a qualitative change of the above picture.

\subsubsection {The case $t_{\text{l}}=1$}
For $t_{\text{l}}=1$ a distinction between two types of vertex defects becomes necessary. Now, unless a new 2-cluster mate is provided by fresh influx during the next iteration, a singleton violates the lower threshold rule. Hence it can only survive for one timestep.

To cope with this situation in the present framework defects consisting of a singleton ($\alpha$-defects) are assigned an energy $\alpha$, defects made of a completely removed 2-cluster pair ($\gamma$-defects) an energy $\gamma$. Both energies are expected to increase with growing bit-chain length $d$ and for higher upper thresholds $t_{\text{u}}$. Then a configurations energy with $l_1$ $\alpha$- and $l_2$ $\gamma$-defects is $E(l_1,l_2)=l_1\alpha+l_2\gamma$.

The probability of having $l_1$ independent $\alpha$- and $l_2$ $\gamma$-defects is
\begin{align}
\label {PPP}
p(l_1,l_2)=\frac{1}{Z}\begin{pmatrix} N/2 \\ l_1 \end{pmatrix}\begin{pmatrix} N/2-l_1 \\l_2\end{pmatrix}2^{l_1}e^{-l_1\beta\alpha}e^{-l_2\beta\gamma}.
\end{align}
For the partition sum one obtains
\begin{align}
\nonumber
Z&=\sum^{N/2}_{l_1=0} \begin{pmatrix} N/2 \\ l_1\end{pmatrix}2^{l_1}e^{-l_1\beta\alpha}\sum^{N/2-l_1}_{l_2=0} \begin{pmatrix} N/2-l_1\\ l_2\end{pmatrix}e^{-l_2\beta\gamma}\\&=(1+2e^{-\beta \alpha}+e^{-\beta \gamma})^{N/2}.
\end{align} 
In analogy with Eq. (\ref{MP1}) one derives the mean number of $\alpha$-defects
\begin{align}
\langle l_1\rangle =N\frac{e^{-\beta \alpha}} {1+2e^{-\beta \alpha}+e^{-\beta \gamma}}
\end{align} and obtains the mean number of $\gamma$-defects
\begin{align}
\langle l_2\rangle=\frac{N}{2} \frac {e^{-\beta\gamma}} {1+e^{-\beta\gamma}+2e^{-\beta\alpha}}.
\end{align}

Finally, the number of empty vertices is $n=N-l_1-2l_2$, and hence the mean population is given by
\begin{align}
\label {MP2}
\langle n\rangle=N \frac{1+e^{-\beta\alpha}} {1+e^{-\beta\gamma}+2e^{-\beta\alpha}}.
\end{align}

\begin{figure}
\begin{center}
  \hspace{0.1cm}
  \epsfxsize 7 cm
  \epsffile {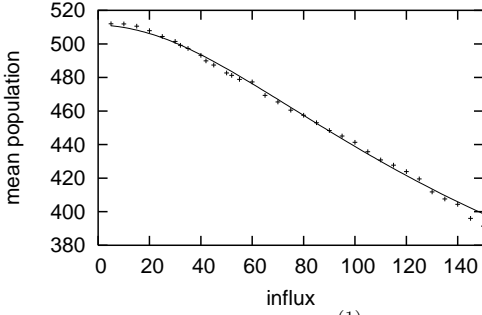}
  \caption {The mean population for $G^{(1)}_{10}$ with ($t_{\text{l}},t_{\text{u}}$)$=$($1,5$). Fitting with Eq. (\ref{MP2}) leads to $\alpha=632\pm99$, $\gamma=250.6\pm13.1$, and $I_0=36.1\pm4.2$. The fit breaks down if one tries to include the range $I \geq 150$ which marks the onset of the transition region (see Sec. \ref{transitionregion}).}
 \label {FIG14}
 \end{center}
 \end{figure}

The ansatz $1/\beta=I_0+I$ is again confirmed by a comparison of Eq. (\ref {MP2}) with numerical data for the mean population (see Fig. \ref{FIG14}).
The probability distribution that $n$ vertices are occupied at a time $t$ is
\begin{align}
\nonumber
\lefteqn{\text{prob} \{ n \text{ occupied vertices}\}=}\\ \label {P2}
 & & \sum^{l_{\text{max}}}_{l_2=0}p(N-n-2l_2, l_2),
\end{align}
where $l_{\text{max}}=\lfloor N/2-n/2\rfloor$ and $p(l_1,l_2)$ as given by Eq. (\ref {PPP}).

\begin{figure}
\begin{center}
  \epsfxsize 7 cm
  \epsffile {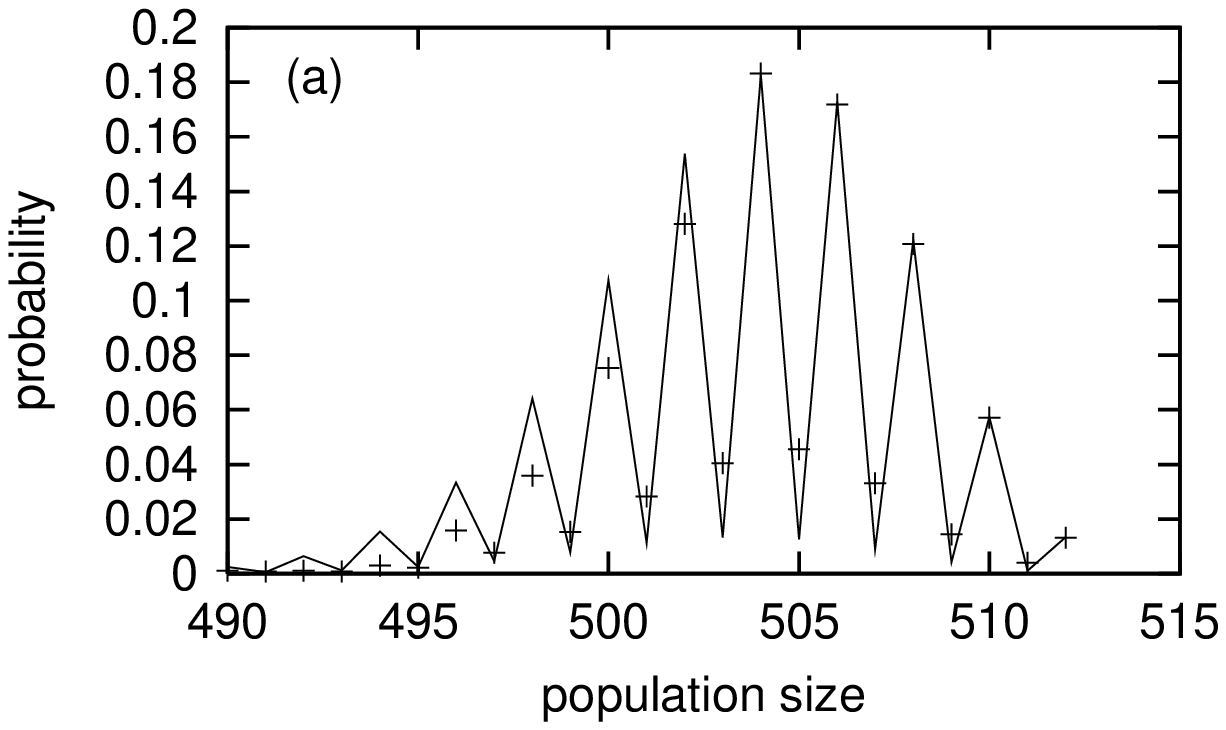}
  \epsfxsize 7 cm
  \epsffile {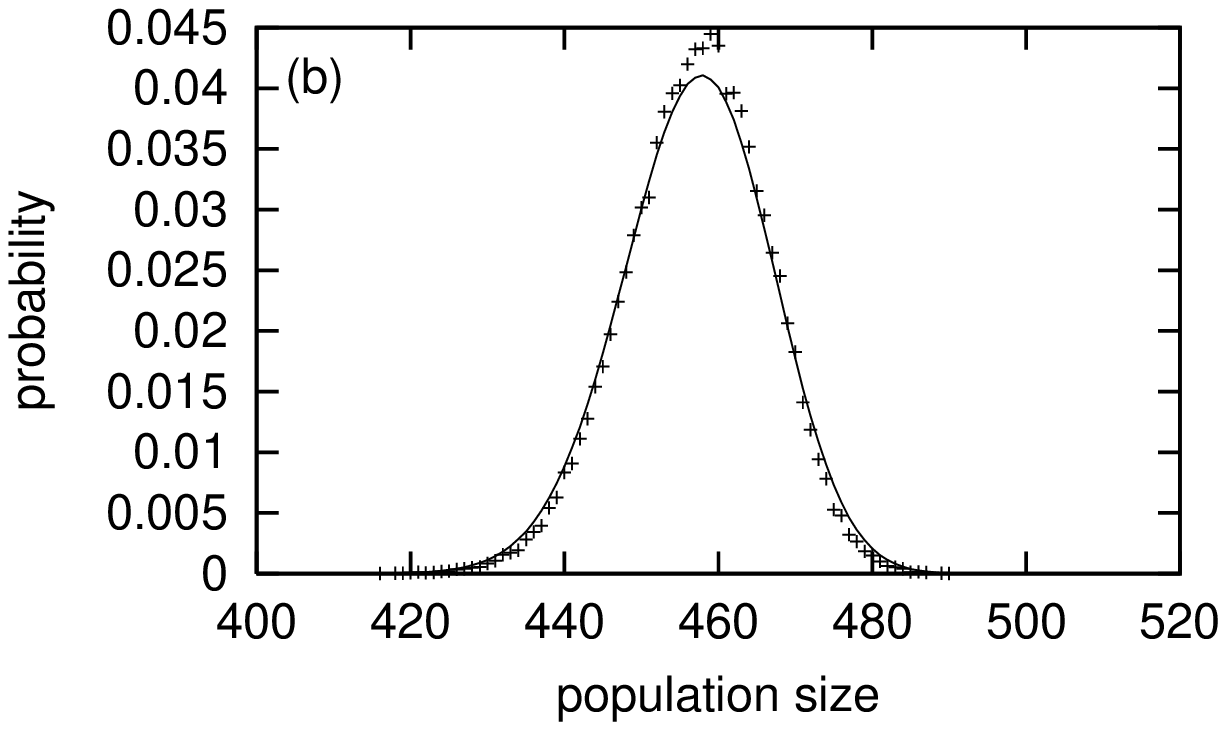}
  \vspace{0.5cm}
  \caption {Comparison between simulation data for the probability distributions of the population size for $G^{(1)}_{10}$, ($t_{\text{l}},t_{\text{u}}$)$=$($1,5$) (a) $I=25$ and (b) $I=80$ and results derived from Eq. (\ref{P2}) for the same $\alpha$, $\gamma$, and $I_0$ as in Fig. \ref{FIG14}. In (b) the influx is already high enough that both types of defects appear with almost equal probabilities.}
 \label {FIG15}
 \end{center}
 \end{figure}

Analytical results derived from Eq. (\ref{P2}) are again in good agreement with simulation data (see Fig. \ref{FIG15}). Equations (\ref{MP2}) and (\ref{P2}) are subject to the same limitations as in the case for $t_{\text{l}}=0$.

\subsection {The transition region}
\label{transitionregion}

As previously discussed increasing $I$ beyond a threshold leads to a growing number of hole defects. The appearance of correlated hole defects is connected with the formation of great clusters consisting of connected 2-clusters. As $I$ becomes still larger, the probability that hole defects become correlated increases. Occasionally this leads to the emergence of a giant cluster (see Fig. \ref{FIG9}, the different series of local maxima correspond to an increasing number of connected hole defects). Speaking in the language of thermodynamics, growing influx entails a growth of fluctuations which finally tend to overthrow the regular pattern structure.
\begin{figure}
\begin{center}
  \epsfxsize 4.1cm \hspace{0.1cm}
   \epsffile{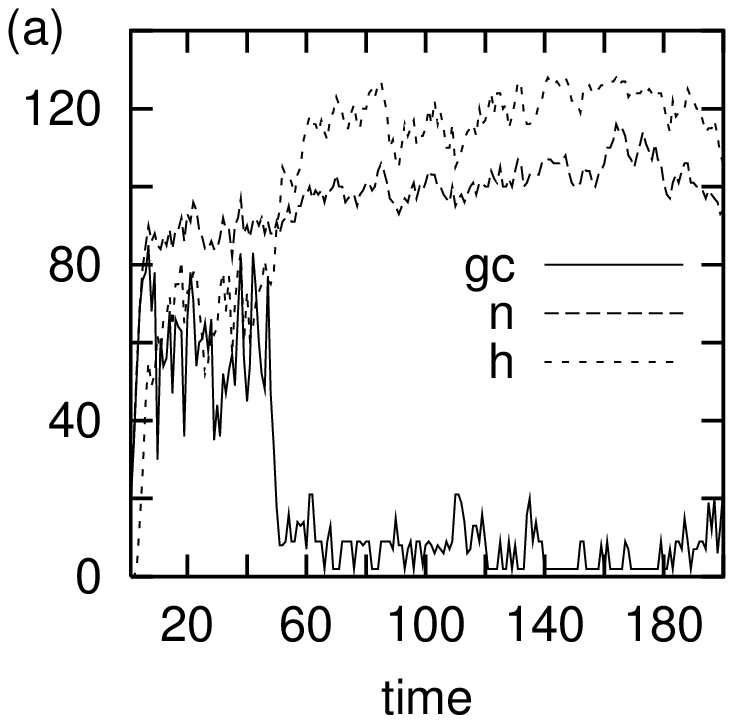} 
   \epsfxsize 4.1cm \epsffile{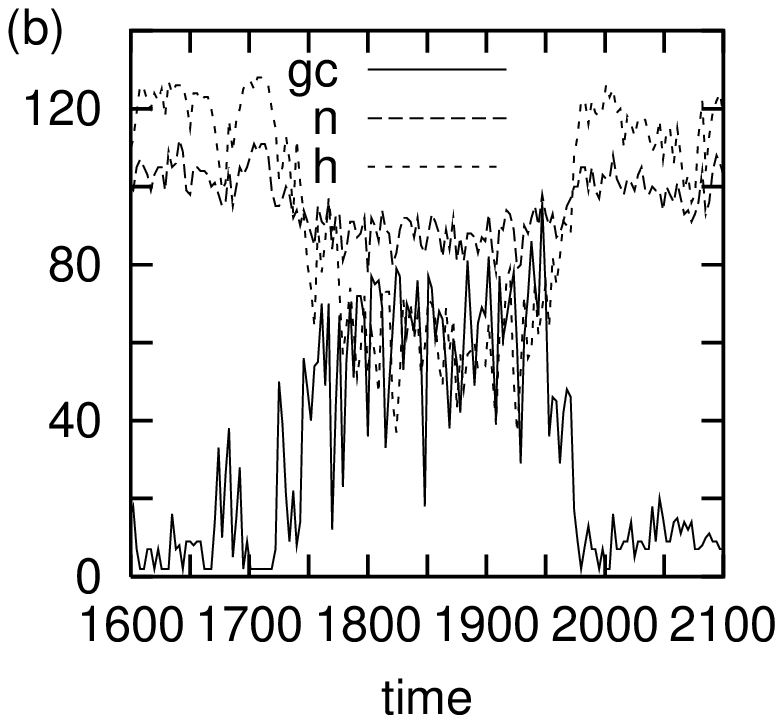} 
   \epsfxsize 4.1cm \hspace {0.1cm} \epsffile{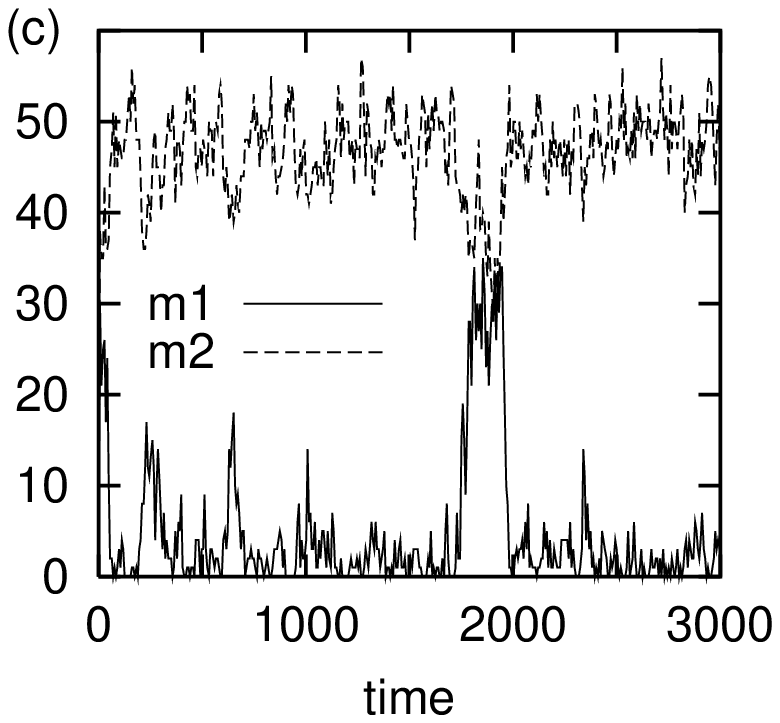} 
   \epsfxsize 4.1cm \epsffile{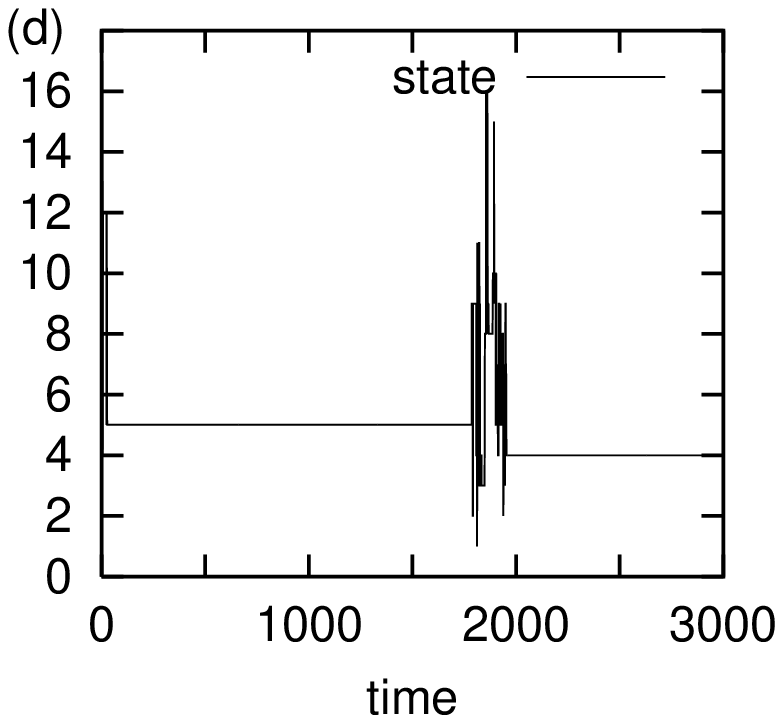}
  \caption {Simulation data for $G^{(1)}_8$, ($t_{\text{l}},t_{\text{u}}$)$=$($1,4$). (a) shows the population $n$, the number $h$ of stable holes, and the size of the greatest connected component of $\Gamma_t$, gc. Initially, approximately half of all sites on $G^{(1)}_8$ become stable holes and a pattern is assumed. (b) shows a transition region. The number of stable holes $h$ and the population $n$ drop suddenly, a giant cluster is formed. (c) shows the smallest and the second smallest distance to a stable base configuration $m_1$ and $m_2$ (cf. text). In the transition region no pattern is selected. (d) displays a number for the pattern which has  the smallest distance to $\Gamma_t$. In the transition region between two sharply assumed patterns the state changes quickly. For $t\in [200,1700]$, $k_{\text{min}}=5$, for $t>2000$, $k_{\text{min}}=4$.}	
  \label {FIG16}
 \end{center}
 \end{figure}

\begin{figure}
\begin{center}
\epsfxsize 4 cm \hspace{0.2cm}
\epsffile{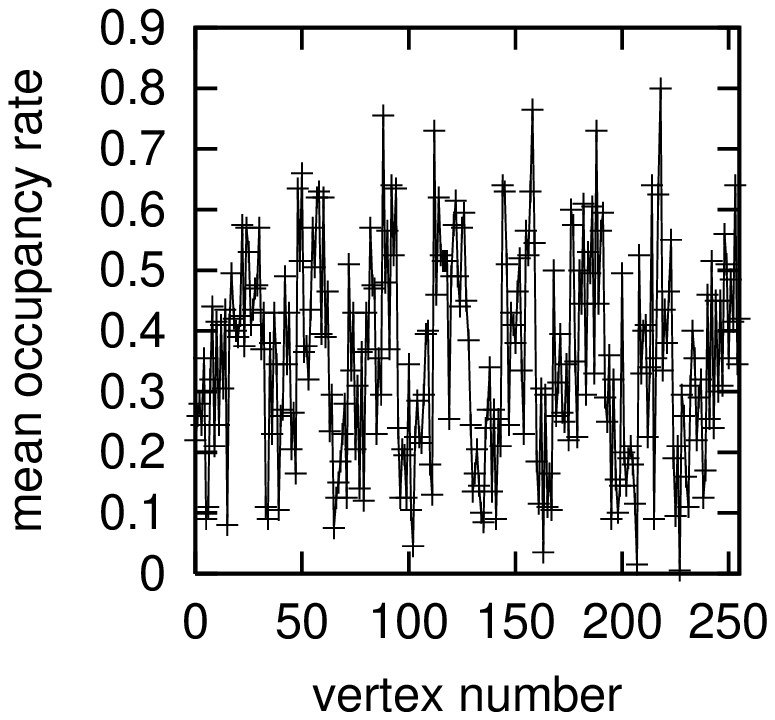}
\epsfxsize 4 cm \hspace{0.2 cm}
\epsffile{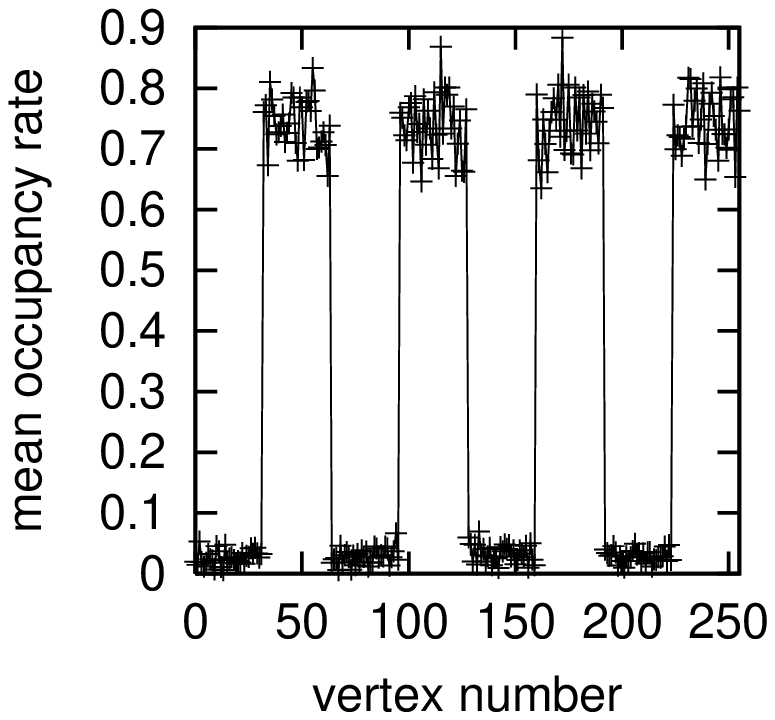}
\vspace{0.5cm}
\caption {Comparison between the mean occupancy distributions measured in a transition region (left) and during the time to the transition region (right). In the right hand figure a 2-clustered pattern is assumed.}
\label {FIG17}
 \end{center}
 \end{figure}

The building of a giant cluster requires the reversion of a major fraction of the stable holes into normal holes. Thus, since stable holes form the `skeleton framing' of the pattern structure, the system relapses into a phase where no pattern lasts for a longer timespan.

Figure \ref{FIG16} illustrates this behaviour for a simulation on $G^{(1)}_8$ with ($t_{\text{l}},t_{\text{u}}$)$=$($1,4$). Initially, after the first stages of the dynamics the system settles into a steady state. For better illustration every stable base configuration $B_{(s_0, P)}$ is given a unique number $k, k=1,...,18$. In Fig. \ref{FIG16}c the minimum distance
\begin{align}
m_1=\min_{k} d(\Gamma_t, B_k)
\end{align} and the second smallest distance
\begin{align}
m_2=\min_{k\neq k_{\text{min}}} d(\Gamma_t, B_k)
\end{align} to any stable base configuration are shown. For the time intervals $[200,1700]$ and $[2000,3000]$ it holds $m_1 \ll m_2$. Consequently the system is in the stable base configuration $B_{k_{\text{min}}}$. Incidentally, hole defects associated with the emergence of larger clusters appear. Mostly,  as long as hole defects remain uncorrelated, they are repaired during the next iterations. At $t\approx 1700$ the exact lock in of state $k_{\text{min}}=5$ is lost and for $t\in I_{\text{change}}=[1700,2000]$ a phase marked by frequent changes of $k_{\text{min}}$ is entered. During this period $m_1\approx m_2$ and thus no base configuration is sharply assumed. A comparison of vertex mean occupancy rates during this phase and the previous one justifies to speak of the time-interval $t\in I_{\text{change}}$ as of a disordered period, since, clearly, any distinction between high mean occupancy and low mean occupancy vertices is lost. As expected, the emergence of the disordered phase goes hand in hand with the formation of a giant connected component in $\Gamma_t$, a drastic loss of stable holes, and a drop in the population.

The whole dynamics of the system is marked by a sequence of ordered and disordered phases. As $I$ is increased, the residence time in ordered phases declines till finally the frequency of `base configuration' changes (defined as a change of $k_{\text{min}}$) tends to one. Thus, the system becomes completely disordered.

\begin{figure}
\begin{center}
\epsfxsize 7 cm
\epsffile {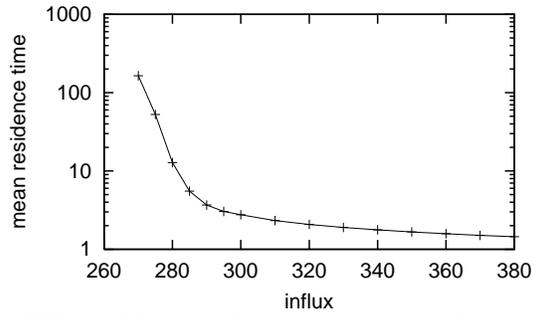}
\caption {Mean residence time, i.e. time during which $k_{\text{min}}$ remains unchanged (excluding a relaxation time of the first 5000 timesteps). All simulations on $G^{(1)}_{10}$ with ($t_{\text{l}},t_{\text{u}}$)$=$($1,8$) have been allowed to run for $10^5$ timesteps. For $I<I_c=270$ no change of $k_{\text{min}}$ has been found.}
\label {FIG18}
 \end{center}
 \end{figure}

\begin{figure}
\begin{center}
\epsfxsize 7 cm
\epsffile{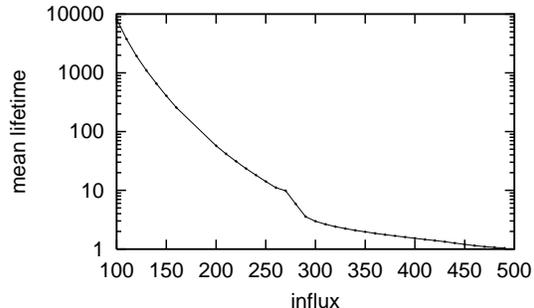}
\caption {Mean lifetime vs. influx $I$ obtained from simulations on $G^{(1)}_{10}$ for ($t_{\text{l}},t_{\text{u}}$)$=$($1,8$). The mean lifetime was determined after allowing $5000$ timesteps for relaxation to a steady state. For every $I$ the simulations then run for 95000 more iterations. Data for $I<100$ are not shown since mean lifetimes exceeded simulation times.}
\label {FIG19}
 \end{center}
 \end{figure}

Figure \ref{FIG18} displays simulation data on $G^{(1)}_{10}$ with ($t_{\text{l}},t_{\text{u}}$)$=$($1,8$) for the mean time (residence time) during which $k_{\text{min}}$ remains unchanged. An abrupt transition from a phase marked by permanent patterns to a phase of transient patterns interrupted by periods of disorder occurs at $I_c\approx 270$. Accordingly, as can be seen from Fig. \ref{FIG19}, the mean lifetime (averaged over all vertices) drops abruptly during this range of influxes. For $I>300$ mean lifetimes decay exponentially and quickly tend to one. Very short lifetimes, however, do not allow for static order or memory. Apparently, the transition from order to disorder goes through a phase of more and more frequent pattern changes.


 \subsection {High $I>I_c$: Iterated maps}
For influxes $I>I_c$ the lifetime of occupied vertices becomes very short. This indicates that the system can no longer form ordered patterns and long-time correlations between the systems configurations disappear. Thus, while for low influx clearly $\langle n_{t+1}\rangle$ is an intricate function of both the systems configuration $\Gamma_t$ at timestep $t$ and the most recent influx, for high influxes the mean population at timestep $t+1$ tends to depend only on the size of $I$ and the mean population at the previous timestep. Also, from the mean occupancy distributions one conjectures, that the structure of $\Gamma_t$ is essentially random, i.e. obtained by randomly occupying $|\Gamma_t|$ vertices of the base graph $G$.

If one randomly occupies $n$ sites of a base graph $G$ the relative mean number $h(n)$ of occupied sites after sites with degrees outside of $(t_{\text{l}},t_{\text{u}})$ have been removed is given by
\begin{align}
\label{iteq}
h(\frac{n}{|G|})= \begin{cases} \frac{n}{|G|} \sum\limits^{t_{\text{u}}}_{l=t_{\text{l}}}\begin{pmatrix} \kappa \\ l\end{pmatrix} (\frac{n}{|G|})^l (1-\frac{n}{|G|})^{\kappa -l} & \text { if $n<|G|$} \\ 0 & \text{ otherwise.}\end{cases}
\end{align}
Considering a large random influx, apart from fluctuations the relative population $x_t$ will evolve according to $x_{t+1}=h(I/|G|+x_t)$. Thus, in the steady state $x$ will preferably assume values near the stable fixed points of the shifted function $g(x)=h(I/|G|+x)$, i.e. near the solutions of $x^\star=g(x^\star)$. Linear stability analysis yields that a fixed point is stable if $|g\prime(x^\star)|<1$. Analogously, $n$-cycles $\{x^\star, g(x^\star), g^{(2)}(x^\star)=g(g(x^\star)), ..., g^{(n-1)}(x^\star) \}$ are solutions of $g^{(n)}(x)=x$ and are stable if $|g^{(n)}\prime (x^\star)|<1$.

For $\kappa=11$ and $(t_{\text{l}}, t_{\text{u}})=(1,8)$ $g(x)$ is a function with one maximum. Consequently, one expects that $g^{(n)}$ has at most $n$ maxima. Thus, in principle, cycles of higher order are possible.

\begin{figure}
\begin{center}
\epsfxsize 9 cm
\epsffile {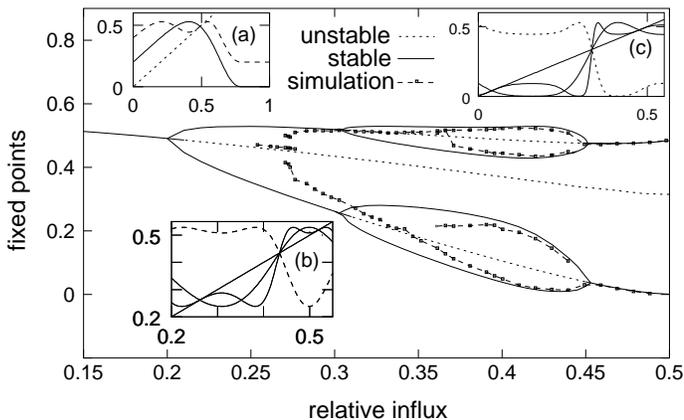}
\vspace {0.5cm}
\caption {The bifuraction diagram obtained by iterating the function $g(x)$ in comparison to simulation results for the maxima of the population histogram for $(t_{\text{l}}, t_{\text{u}})=(1,8)$ on $G^{(1)}_{10}$ (filled squares connected by lines). Solid lines correspond to stable branches, dashed lines to unstable branches. The region of the simulation data where three stable solutions coexist corresponds to the onset of disorder: The intermediate branch represents the mean population during periods in which the system closely assumes a base configuration, the upper and the lower branch are data sampled during disordered phases. Especially for higher influx simulation results and the data obtained by iterating $g$ are in very good agreement. The smaller diagrams visualize changes in the stability of fixed points of $g$ at the branching points: (a) $g$ (solid line) becomes unstable at $I/|G|=0.2$, but $g^{(2)}$ (dashed line) is stable. Hence a cycle appears. (b) At $I/|G|=0.3$ also $g^{(2)}$ (solid line) becomes unstable and --since $g^{(3)}$ (dashed line) gives only one unstable solution-- a 4-cycle emerges. (c) At $I/|G|=0.45$ $g^{(2)}$ becomes stable again leading to a backward bifurcation and a 2-cycle.}
\label{FIG19_1}
\end{center}
\end{figure}

The function $g(x)$ is a polynomial of order $t_{\text{u}}$. Thus $g^{(n)}$ is of order $t_{\text{u}}^n$ and there is no general analytic solution to $x^\star=g^{(n)}(x^\star)$. As an example we consider the case $(t_{\text{l}}, t_{\text{u}})=(1,8)$ on $G^{(1)}_{10}$, i.e. $\kappa =11$ and $|G|=1024$. Evaluating the fixed point equations numerically, one obtains the bifurcation diagram displayed in Fig. \ref{FIG19_1}. At $x=0.2$ one finds a forward pitchfork bifurcation towards a 2-cycle, which then at $x=0.3$ becomes also unstable, leading to the emergence of a 4-cycle. For $x=0.45$ this 4-cycle is found to become unstable and a backward pitchfork bifurcations leads again to a stable 2-cycle. Far outside the region of biological interest, for $x=0.93$ a backward bifurcation occurs and the single fixed point solution again becomes stable. Figure \ref{FIG19_1} also displays data obtained from simulations of the window algorithm on $G^{(1)}_{10}$ which exhibit the same qualitative behaviour as the iteration of $g$. As expected, deviations become small as $I$ increases. 

\section {More densely wired systems: two-mismatch base graphs }
In this Section some observations about the dynamics created by the window algorithm with $t_{\text{l}}=1$ on more densely wired base graphs will be presented. As a consequence of a higher coordination number different static patterns and a new type of dynamic pattern can be observed. In principle, we expect (i) more sparsely populated, and hence transient patterns of reduced stability for low $t_{\text{u}}$ and (ii) enhanced stability for upper thresholds close to $\kappa-1$ (cf. Sec. \ref{patternstability}).

As an extension of the base graphs $G^{(1)}_d$ whose link structure is created by `one-mismatch links' we now introduce additional links connecting bit-chains which deviate in two positions from complementarity, so defining the base graphs $G^{(2)}_d$, cf. Eq. (\ref{mismatcheq}).

Clearly, since $G^{(1)}_d$ is connected such `two-mismatch links' of $G^{(2)}_d$ are shortcuts of paths in $G^{(1)}_d$. For example, referring to the notion of bit-operations associated with links (see Sec. \ref{modelinginws}), two vertices linked by $L_{k_1 k_2}\in G^{(2)}_d$, $k_1\neq k_2$, are also linked by the six paths corresponding to combinations of the edges $L_0$, $L_{k_1}$, $L_{k_2}\in G^{(1)}_d$.

\begin{figure}
\begin{center}
\epsfxsize 4 cm
\hspace{0.3cm}
\epsffile {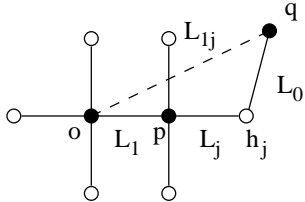}
\vspace{0.5cm}
\caption {An illustration of how `two-mismatch links' (dashed line) contribute to the degree of occupied vertices in a base configuration $B_{(s_0, L_1)}$. Occupied vertices in $B_{(s_0, L_1)}$ are drawn in black, empty ones in white. Two-clusters (e.g. $o$ and $p$) in $B$ are connected via the `one-mismatch link' (solid line) $L_1$. Except $o$ the one mismatch neighbours $h_j=L_j (p)$, $j\neq 1$, of $p$ are holes. Apart from the holes' 2-cluster mates $L_1(h_j)$ all their neighbours are occupied.  If one pursues an inversion link $L_0$ from any such hole $h_j$ adjacent to $p$, a vertex $q=F_0(h_j)$ is reached which is --via $L_{1 j}$-- also a two mismatch neighbour of $o$. One can easily verify that $o$ has no other occupied `two mismatch neighbours'. Consequently, in a base configuration $B$ on $G^{(2)}_d$ every occupied vertex has $d+1$ occupied neighbours.}
\label {FIG20}
 \end{center}
\end{figure}

Although every vertex now acquires $d(d-1)/2$ new edges the former link structure is conserved. Thus, it could be conceived that patterns created on $G^{(2)}_d$ retain a part of the structure of the stable base configurations which are formed on $G^{(1)}_d$. Counting, e.g., `two-mismatch' links in a base configuration $B_{(s_0, F_{1})}\subset G^{(1)}_d$ one finds that every occupied vertex now acquires $d$ new links on $G^{(2)}_d$ (cf. Fig. \ref{FIG20}), aggregating to a total degree $d+1$. Thus, this kind of patterns is forbidden for $t_{\text{u}}<d+1$. However, it turns out that apart from metastable structures for small influx the dynamics always ends up in the above described configurations. Thus, for $t_{\text{u}}\geq d+1$ the base-graph $G^{(2)}_d$ always decays into two groups of high mean occupancy vertices $S_H$ and low mean occupancy vertices $S_L$. Both, $S_H$ and $S_L$ form $(d+1)$-regular subclusters of $G^{(2)}_d$. Resuming the considerations of Sec. \ref{statedescription} one finds that these stable base configurations on $G^{(2)}_d$ have multiplicity $2d$. The scenario for growing influx is the same as on $G^{(1)}_d$: for $I>I_c$ phases of well-defined base configuration are interrupted by disordered periods leading to pattern-changes, and finally to a dynamics marked by randomness and very short lifetimes.

However, for $t_{\text{u}}<d+1$ where `relics' of patterns forming on $G^{(1)}_d$ are no longer allowed the behaviour turns out to be more interesting.


\begin{figure}
\begin{center}
\epsfxsize 7 cm
\hspace{0.3cm}
\epsffile {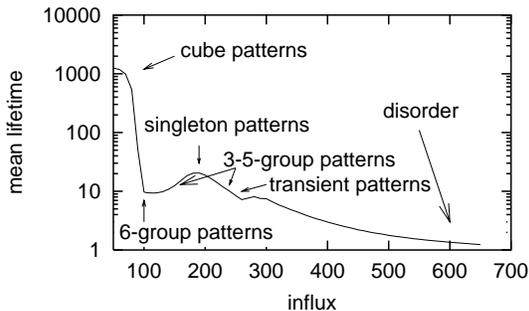}
\vspace{0.5cm}
\caption {Mean lifetime (log-scale) vs. influx for simulations on $G^{(2)}_{12}$ with ($t_{\text{l}},t_{\text{u}}$)$=$($1,10$). Successively, the system forms cube-patterns, multiple-group patterns, singleton patterns, and again multiple-group patterns. Then, patterns become transient and finally a randomness-driven regime is entered (see text).}
\label {FIG21}
 \end{center}
\end{figure}

As tools to investigate the structure of networks we again consider the mean occupancy rates of all vertices and probe the mean occupancy-patterns by defining subsets $S(a)\subset G^{(2)}_d$ with different threshold-occupancy rates $a$ (see Sec. \ref{thestationarystate}). In analogy with results for the one-mismatch case we find metastable patterns for small influxes. Depending on the ratio $I/|G^{(2)}_d|$, first completely 2-clustered, 4-clustered or 8-clustered patterns appear (4-clusters are chains, 8-clusters cubes). Next in a range of influxes cube-configurations dominate (cf. Fig. \ref{FIG21}). Different from the 2- and 4-cluster patterns, however, the cubes don't fill the base-graph completely and 256 vertices not matching in the cube pattern are left.

\begin{figure}
\begin{center}
\epsfxsize 7 cm
\hspace{0.3cm}
\epsffile {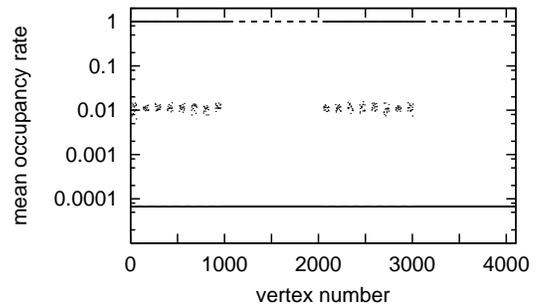}
\vspace{0.5cm}
\caption {Every vertex' mean occupancy obtained by a simulations on $G^{(2)}_{12}$ with ($t_{\text{l}},t_{\text{u}}$)$=$($1,10$) and $I=30$. Vertices are labelled by integers z corresponding to their respective bit-chains. The relaxation time was $T_0=5000$, mean occupancies have been sampled over $T_1-T_0=15000$ further timesteps. An offset has been added to display the group $S_L$ with the lowest mean occupancy $\langle a\rangle_{S_L}=0$. The intermediate group $S_M$ has mean occupancy $\langle a\rangle_{S_M}\approx 0.01$, the group with highest mean occupancy $\langle a\rangle_{S_H}\approx 1.0$.}
\label {FIG22}
 \end{center}
\end{figure} Figure \ref{FIG22} shows data for every vertex' mean occupancy obtained by a simulation on $G^{(2)}_{12}$ with $(t_{\text{l}}, t_{\text{u}})=(1, 10)$. The set of vertices with highest mean occupancy $S_H$ decays into cubes, the set with lowest mean occupancy $S_L$ forms a giant component of $2816$ vertices. $S_L$ consists of 1536 almost never occupied vertices of degree $\partial_{|G\backslash S_L} =61$, 256 with $\partial_{|G\backslash S_L} =43$ and 1024 with $\partial_{|G\backslash S_L} =39$. All degrees are larger than $t_{\text{u}}$ and thus $S_L$ entirely consists of stable holes. The 256 vertices not exactly matching in the cube pattern form a subset $S_M$ of intermediate mean occupancy. $S_M$ turns out to contain only singletons, each of which is surrounded by elements of $S_L$ which are almost never occupied. Thus, vertices of $S_M$ are isolated spots of activity sustained by influx temporarily placed at their surrounding stable holes, but taken out immediately afterwards due to overstimulation. Consequently, as long as no fresh influx is placed adjacent to it in the next timestep a vertex of $S_M$ dies out one timestep after it has been inserted. This view is confirmed by very short mean lifetimes $\langle T_{\text{life}}\rangle_{S_M}\approx 1$ of elements of $S_M$, which slightly grow with increasing influx. Together with a growing probability that isolated holes get filled this accounts for relatively large mean occupancies  $\langle a\rangle_{S_M}$ in comparison to $\langle a\rangle_{S_L}$. Though at the first glance surprising that a vertex can survive at a site where it is surrounded by almost always empty sites, this but indicates a mechanism which is to prevail in a consecutive regime: isolated vertices of high durability are sustained by quickly fluctuating short-living influx in their neighbourhoods.

\begin{figure}[t!]
\begin{center}
\epsfxsize 7 cm
\hspace{0.3cm}
\epsffile {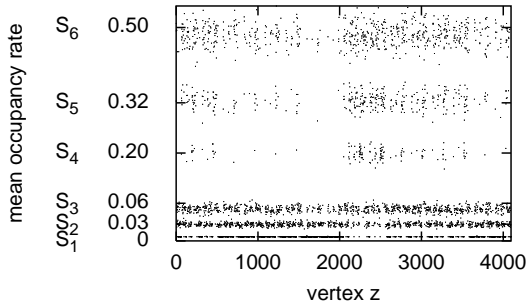}
\vspace{0.5cm}
\caption {Mean occupancy rates of vertices $z$ derived from a simulation on $G^{(2)}_{12}$ with ($t_{\text{l}}, t_{\text{u}}$)$=$($1,10$) and $I=100$. Six groups of vertices $S_1,...,S_6\subset G^{(2)}_{12}$ can be distinguished by their mean occupancy.}
\label {FIG22a}
 \end{center}
\end{figure}

Increasing $I$ over approximately 80 the cube patterns become unstable and from $I\approx 100$ onwards a pattern marked by 6-levels of mean occupancy (see Fig. \ref{FIG22a}) is chosen. In contrast to the appearance of several levels of mean occupancy due to transitions between isomorphic stable base configurations this grouping is of a different nature. While in the first case simulations started with different initial conditions lead to different numbers of groups and changing mean occupancy levels (depending on the timestep when transitions occur) simulations performed with different initial conditions now always reproduce these 6 groups and their mean occupancies.

\begin{table}[b]
\begin{center}
\begin{tabular} {c c c c c c c c}
 group $S$& $|S|$ & $\langle C\rangle$ & $\langle \tau \rangle$  & $\langle r \rangle$ & $\langle f_{\text{small}}\rangle$ & $\langle C\rangle_{\text{small}}$ & $\langle n\rangle$ \\ \hline \\[-0.3cm]
$S_1$ & 1124 & $371.0$ & $0.0$ & $1.0$ & 0.0 & - & 0.05 \\
$S_2$ & 924 & $371.0$ & $3.8$ & $0.26$ & 0.0 & 7.0 & $27.2$ \\
$S_3$ & 924 & $371.0$ & $5.4$ & $0.18$ & 0.0 & 4.6 & $61.3$ \\
$S_4$ & 134 & $1.0$   & $10.0$ & $0.1$ & 1.0 & 1.0 & $26.5$ \\
$S_5$ & 330 & $160$   & $18.1$ & 0.06  & 0.5 & 1.007 & 107.05 \\
$S_6$ & 660 & $260$   & $35.3$ & 0.028 & 0.27& 1.03 & 318.16 \\
\end{tabular}
\vspace {0.5cm}
\caption {Data characterizing the 6 groups $S_1$,$...$, $S_6$ obtained from simulations on $G^{(2)}_{12}$ with ($t_{\text{l}}, t_{\text{u}}$)$=$($1,10$) and $I=100$. In 30000 timesteps vertices of $S_1$ have always belonged to the greatest subcluster of $\Gamma_t$, hence $\langle C\rangle_{\text{small}}$ could not be measured. Since they are immediately taken out after they got into the system stable holes have a mean lifetime $\langle \tau \rangle =0.0$. This means that there mean lifetime is shorter than the discretization of time. }
\label{TAB1}
\end{center}
\end{table}
The interval of influxes $I\in [80,100]$ is characterized by transitions between both types of patterns in the course of which for higher influxes periods during wich the 6-group pattern is assumed more and more prevail. The change of the mean population averaged in both types of pattern is drastic: for $I=90$ the cube-pattern can sustain approximately $\langle n\rangle\approx 1024$ vertices while at the onset of their appearance 6-groups patterns can support a mean population of only about $\langle n\rangle\approx 600$ occupied sites. Comparing the mean occupancy levels of the 6-group pattern with all previous pattern structures one remarks that, while groups of always empty sites are still present, groups of almost permanently occupied vertices are lacking. All groups have mean occupancy rates smaller than $0.7$. This, and the relatively short lifetimes of vertices belonging to the groups $S_4,S_5$, and $S_6$ with high mean occupancy, speaks in favour of an interpretation as a `dynamic' pattern. Occupied vertices belonging to different groups `fluctuate' with different rates, while the interplay of them secures that the distinct role of the groups is preserved. In the following the nature of this dynamic pattern will be elucidated in more detail.

Figure \ref{FIG22a} and Table \ref{TAB1} give data characterizing this six groups: mean size of the cluster $\langle C\rangle$ on $\Gamma_t$ that a vertex of a group belongs to, its mean lifetime $\langle \tau \rangle$, mean switch rate $\langle r \rangle$, the mean frequency $\langle f_{\text{small}}\rangle$ of belonging not to the giant component and the mean cluster size $\langle C\rangle_{\text{small}}$ of the small clusters if the vertex is not in the greatest component of $\Gamma_t$. On average elements of $S_1, S_2$, and $S_3$ belong to $371$-clusters on $\Gamma_t$, but are very seldom members of small clusters or singletons, have relatively short lifetimes and high switch rates. In contrast, elements of the groups of higher mean occupancy $S_4,S_5$, and $S_6$ typically have smaller mean cluster-sizes, form more frequently small clusters, preferably singletons, and have long lifetimes. Mean lifetimes are increasing with increasing mean occupancy of the groups.

\begin{figure}
\begin{center}
\epsfxsize 6 cm
\hspace{0.3cm}
\epsffile {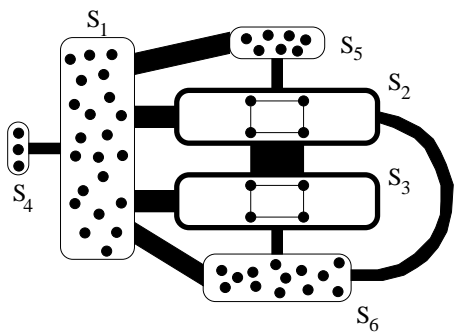}
\vspace{0.5cm}
\caption {Visualization of the network structure of the groups $S_1$,...,$S_6$. The squares indicate many connections of vertices of a group among themselves, isolated circles in a group visualize that it consists of singletons. The thickness of the connecting lines gives a measure of the number of links connecting elements of different groups. Because of their importance for the existence of giant components in $\Gamma_t$ the core groups $S_2$ and $S_3$ are bounded by fat lines. Larger boxes correspond to larger groups.}
\label{FIG23}
 \end{center}
\end{figure}

Exploring the interconnectedness and network structure of the groups $S_1,...,S_6$ one obtains the following picture (cf. Fig. \ref{FIG23}): The strongly interconnected groups $S_2$ and $S_3$ form a core of $G$ to which --except $S_4$-- all the other groups are attached. Excluding this core, each of the other groups consists of singletons. The largest set $S_1$ is a reservoire of stable holes and connects $S_4$ with the core. Thus, vertices of $S_4$ play a similar role as the previously mentioned isolated holes in the cube-pattern: they receive their whole sustenance from the influx and have no permanent connections. While the group of second largest mean occupancy, $S_5$, has only links to the less populated part of the core, $S_2$; the most active vertices, elements of $S_6$, have connections with both parts of the core.

Consequently, apart from stimulation by random influx elements of $S_5$ obtain stimulus from the core-group $S_3$. Nevertheless, the stimulation is not optimal leading to a mean occupation of only $1/3$ of $S_5$. The group $S_6$ receives stimulation from both core groups and so can support a higher population than $S_5$. Contrarily, since being strongly interconnected and connected to both the richely populated groups $S_5$ and $S_6$ most vertices of $S_2$ are suppressed. In the case of $S_3$ which does not connect to $S_5$ suppression is lower and thus a higher mean population can be sustained. The data shown in Table \ref{TAB1} are in qualitative agreement with this interpretation.

\begin{figure}
\begin{center}
\epsfxsize 7 cm
\hspace{0.3cm}
\epsffile {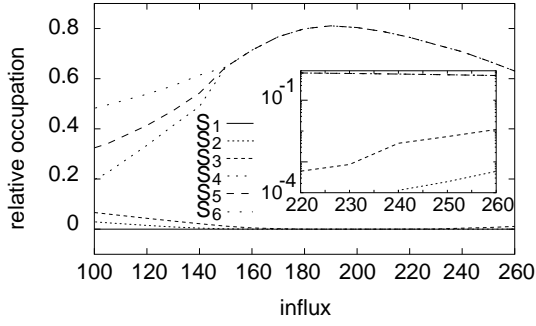}
\vspace{0.5cm}
\caption {Relative occupation $s_i=1/|S_i|\sum_{v\in S_i}\bar{s}(v)$ of the 6 groups (labelled by their group number) for different influxes. For $I\approx 140$ groups $S_1$ and $S_2$ unite, for $I\approx 175$ $S_3$ joins them. At $I\approx 220$ $S_3$ again becomes separated, for $I\approx 240$ also $S_2$ disjoins from $S_1$. Similarly, $S_4$, $S_5$ and $S_6$ merge at $I\approx 150$. This leads to 5-group patterns ($I\in [140,150]$), 4-group patterns ($I\in [150, 175]$ and $I\in [240,260]$), singleton-patterns ($I\in [175,220]$) and a 3-group pattern ($I\in [220,240]$). For $I>260$ transitions between isomorphic configurations set in.}
\label {FIG24}
 \end{center}
\end{figure}

Figure \ref{FIG24} displays simulation data for the change of the relative occupations $s_i=1/|S_i|\sum_{v\in S_i}\bar{s}(v)$, $i=1,...,6$, of the 6 groups with increasing influx. Initially, all groups are clearly distinct, groups 4-6 have relatively low occupation and the mean populations of $S_2$ and $S_3$ are clearly above zero. As $I$ grows while $S_2$ and $S_3$ lose  population $S_4$, $S_5$ and $S_6$ become more populated, but at different rates. Thus, at $I\approx 150$ the groups $S_4$, $S_5$ and $S_6$ become united leading to the formation of a single set $S_H$ of high mean occupancy vertices. Similarly, in the case of $S_1, S_2$, and $S_3$. For $I\approx 140$ group $S_2$ joins $S_1$ and becomes indistinctive from $S_3$ at $I\approx 175$ leading thus to a single low occupancy group $S_L$. During this process the overall mean population of $\Gamma_t$, $\langle n\rangle$, is growing. Consequently, for $I\in [175,220]$ again a 2-group pattern of $1124$ high mean occupancy and $2972$ low mean occupancy vertices appears, consisting now entirely of long living singletons `sustained by the influx'. Transition patterns before this pattern appears are $5$- and $4$-group patterns, after it grew unstable $3$-group patterns are forming.

As $I$ is further increased the influx which heretofore guaranteed the sustenance of $S_5$ and $S_6$ gradually causes overstimulation. Thus, the mean populations of $S_5$ and $S_6$ are shrinking, causing in turn less overstimulation to the core. So, again a small population can persist in $S_2$ and $S_3$. Altogether, for further increased influx the different groups disentangle, leading to a revival of the multi-group patterns.

From $I\approx 260$ on these dynamic multiple group patterns become unstable and rearrangements between different (isomorphic) configurations set in. Analogous to the scenario discussed in Sec. \ref{transitionregion} disordered periods more and more dominate and finally lead to disorder.

Interestingly, the group structure allows some insight into the actual cluster structure of $\Gamma_t$ at an arbitrary timestep. Obviously, giant and greater clusters on $\Gamma_t$ can only emerge as long as the core is populated. As the mean population of $S_2$ does not suffice to form a giant component within $S_2$, $S_3$ functions as the `glue' of giant clusters on $\Gamma_t$. Elements of the `backbone' of the giant cluster preferably come from $S_2$ and $S_3$. As the population of $S_2$ decreases vertices of $S_5$ increasingly become detached from the core and tend to form small clusters and singletons.

A great component, however, is still retained by elements of $S_6$ and $S_2$ bound together by $S_3$. As the population of $S_2$ also sinks below a certain threshold, greater components of $\Gamma_t$ become star-like, their hubs being in $S_3$. Finally clusters different from singletons completely vanish, resulting in purely influx sustained patterns of singletons.

\section {Patterns for {\large \lowercase{$t_{\text{l}}>1$}}}
For completeness in this section stationary patterns appearing for lower thresholds $t_{\text{l}}>1$ will be sketched briefly.

\begin{figure}
\begin{center}
\epsfxsize 7 cm
\hspace{0.3cm}
\epsffile {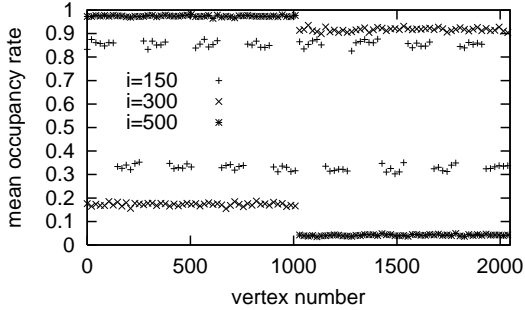}
\vspace{0.5cm}
\caption {The mean occupancy rate of every vertex from simulations on $G^{(1)}_{11}$ with ($t_{\text{l}},t_{\text{u}}$)$=$($3,10$) and influxes $I=150$, $300$ and $500$. For low influx both high-mean-occupancy group and low-mean-occupancy group are considerably apart from their extreme values $1$ and $0$. As $I$ grows, $S_L$ becomes less occupied and vertices of $S_H$ evolve to almost permanent presence. Thus, increasing $I$ further the high mean occupancy set loses stimulation and the pattern becomes unstable. }
\label {FIG25}
 \end{center}
\end{figure}

Figure \ref{FIG25} displays data for the mean occupancy rates of every vertex gathered from a simulation on $G^{(1)}_{11}$ with ($t_{\text{l}},t_{\text{u}}$)$=$($3,10$). One realizes two relatively broad bands of mean occupancy rates. Investigating the structures of $S_H=S(0.5)$ and $S_L=G\backslash S_H$ reveals that, analogously to the $t_{\text{l}}\leq1$-situation both high- and low-mean occupancy groups completely decay into 2-clusters. Now, however, such 2-cluster patterns are only existing with mean occupancy rates $\langle a\rangle_{S_H}$ significantly smaller than $1$ and $\langle a\rangle_{S_L}$ far above zero. Thus, since a `frozen' 2-clustered state (see Sec. \ref{analyzingthedynamics}) can not subsist for $t_{\text{l}}>1$, a fluctuating 2-cluster pattern is formed. At every single timestep only a part of the high-mean occupancy vertices is occupied. As a consequence, not all holes are stable. A part of the holes can become occupied, thus giving the necessary stimulation to the vertices of $S_H$ which are present. On average, whereas elements of $S_L$ have high degrees (and are on the verge of being overstimulated) elements of $S_H$ have low degrees (and are threatened by understimulation). On the whole, however, vertices occupying holes tend to get removed with higher probability. One can imagine this situation as a pattern of long living occupied 2-clusters flown round by short living 1- and 2-clusters, living just for their stimulation.

Considering this situation as evolved from an ideal 2-cluster pattern $B_{(s_0, P)}$ one understands that it can only survive for an intermediate range of influxes. On the one hand, if the influx is too low, not enough occupied sites are driven out by overstimulation that holes could lose stability. Thus, the sites occupied in the ideal pattern $B$ did not receive the necessary stimulation and the pattern was unstable. On the other hand, too high influx causes more and more of the frequently occupied vertices to be almost permanently occupied. This but leads to many stable holes. So, for higher influxes an abrupt transition to disorder occurs. The intermediate behaviour-- transitions between isomorphic configurations-- is lacking in this case.

\section {Conclusions}
To conclude, we have presented a probabilistic model for a local-rule governed evolution of occupied and empty sites on regular graphs $G$. Aiming chiefly at a description of INW's in the immune system we studied two types of base graphs $G^{(1)}$ and $G^{(2)}$ created by describing idiotypes by bit-strings and their functional interactions by `matching-rules'. In contrast to most modelling approaches to idiotypic networks \cite{Ueb}, the model abstains from all details of the dynamics of the real-life interactions of cells, but aims at understanding principal mechanisms of network formation.

On $G^{(1)}$ and $G^{(2)}$ the dynamics generated by the window algorithm leads to organized network structures, which consist of functionally different subsets, distinguished, e.g., by their mean occupancy. On both types of base graphs for influxes $I$ small compared to the system size, a multitude of such network-patterns has been found. We developed a notion of pattern-stability and classified them as metastable. For larger influxes one type of pattern prevails. In the case of the one-mismatch graphs $G^{(1)}$ patterns for intermediate influx always consist of an arrangement of 2-clusters of holes and occupied vertices. We have given a classification of these patterns and have described fluctuations around an ideal pattern structure by a statistical approach via defects. For $I>I_c$ these 2-clustered patterns become unstable, and the system starts oscillating between periods during which a pattern structure prevails and periods of disorder. Finally, increasing $I$ further, the dynamics is marked by randomness and extremely short lifetimes. We elucidated the nature of these transitions.

These 2-clusters are nothing but the idiotype-antiidiotype pairs which have been proposed as one mechanism for the preservation of memory of previously encountered antigen (The antiidiotype represents the internal image of the antigen \cite{Jerne,Behn2}). Their arrangement in a coherent pattern, however, leads to an overall memory capacity of the network which increases only logarithmically with the system size. This is clearly insufficient for a real immune system. The parameter range where the coherent 2-cluster patterns emerge can therefore not be the working regime of a healthy immune system.

It is worthwhile to note that in this regime (and in all regimes allowing for memory on $G^{(2)}$) the network connectivity decreases with growing simulation time, i.e. with growing age. This is in accord with experimental observations and was also found in other modelling approaches \cite{Ueb}. It suggests that the networks gradually losing links is an essence of the limited range of allowed vertex-degrees and hence of cross-linking.

On the more densely wired graphs $G^{(2)}$ cube patterns have been found for an intermediate range of influxes. Increasing $I$ beyond a threshold, the system then settles into dynamic `multiple-group' patterns. In this regime, several subsets $S_j$ of the base graph $G^{(2)}$ can be distinguished according to their mean occupancy. Vertices of these subsets also differ in the structure of the cluster on $\Gamma_t$ which they typically belong to. Two of these subsets form a strongly interconnected core on $\Gamma_t$, to which vertices of the other groups are bound with different strengthes. Each of the other subsets consists only of singletons and links to the core and some of the other groups. The core groups have been found to be always rarely populated, the major fraction of the population being typically contained in two of the other groups.

The structure encountered in this parameter regime is in very good agreement with general ideas about the topology of INW's \cite{Jerne,Behn1,Coutinho,Cohen,Varela}. It exhibits a structured core which could correspond to a central part generally believed to exist in INW's. Furthermore, richely populated non-core groups are in good agreement with the notion of a peripheral part of INW's. Long lifetimes of vertices belonging to these groups are in accord with the idea that the peripheral part of the network is responsible for the preservation of idiotypic memory.

The issue about the in detail-working of idiotypic memory is still much disputed. Apart from the `internal image hypothesis' several mechnisms are currently known \cite{Zinkernagel,Sprent,Freitas,Nayak}. One of these bases on certain cells (folicular dendritic cells) which occasionally exhibit parts of antigen, which is deemed enough to sufficiently stimulate the complementary idiotype -- allowing in this way idiotypic memory in a single antigen-specific clone.

Derived from our model a new mechanism for idiotypic memory seems possible: Memory could be retained in a single antigen-specific clone which is not preferably stimulated, but occasionally receives stimulation from the central part of the network and from new complementary idiotypes from the bone marrow which serves as a background stimulus. This mechanism could collaborate with antigen presentation by follicular dendritic cells.

Two further issues which follow from the multi-group structure of networks in our model seem worthwhile to note.

First, during its evolution the system generates a group $S_1$ of always suppressed idiotypes. Although the dynamics of antigens differs from that of idiotypes, the existence of such a group implies that there should be a set of antigens against which an individual is immune without ever having been immunized against. The actual location of this set is determined by encounters with other antigens and, chiefly, by the fortuitous history of the deployment of bone marrow influx during early life.

As a second fact, while the number of autoantibodies increases \cite {Attanasio} the bone marrow production is known to decrease with growing age \cite{Freitas}. Here, additional to a decline in connectivity till stationarity is reached, our model suggests another structural alteration of the network. Assuming for young individuals a working point of INW's which is above the parameter regime of single-clone patterns, a decreased bone marrow production leads to smaller cores and an increasing periphery, i.e. an accumulation of memory and loss of plasticity.

Summarizing our main result, taking into account different levels of coarse-graining of reaction affinities between idiotypes --as represented by the base graphs $G^{(1)}_d$ and $G^{(2)}_d$-- leads to two different regimes of steady-state behaviours: (i) rigid `coherent' configurations on $G^{(1)}$ and (ii) dynamic patterns on $G^{(2)}$ which are  increasingly dominated by 1-clones. We conjecture that generally isolated clones are a consequence of a high connectivity of the network and high interaction strength between idiotypes. It is suggestive that the bone marrow influx is a driving mechanism for shaping the structure of INW's, which together with the interaction strength of idiotypes, determines the working point of an INW.
\begin{acknowledgements}
M.B. gratefully acknowledges financial support by the S\"achsische Graduiertenf\"orderung.
\end{acknowledgements}


\begin{thebibliography}{99}
\bibitem{FN0} {For $t_{\text{l}}=0$ the window algorithm can be re-formulated from a holes' point of view. Let $\partial_0v=\kappa-\partial v$ denote the number of empty sites adjacent to a vertex $v$. According to step (2) of the window algorithm an occupied vertex is taken out if $\partial v>t_{\text{u}}$. The idea of stable holes implies that holes $h$ can only be taken out if $\partial h=\kappa-\partial_0 h<t_{\text{u}}+1$. Thus, holes $h$ can only be removed if $\partial_0 h >t_{\text{u}}^{\text{hole}}=\kappa-t_{\text{u}}+1$, which is --considering timescales at which holes are probed frequently enough-- the same rule as for occupied vertices.}
\bibitem{FN1a} {There are $\begin{pmatrix} I \\ j \end{pmatrix}$ variants to choose $j$ out of $I$ vertices which are to be placed in the neighbourhood of a specified occupied vertex. Then, since it has a 2-cluster mate, $\begin{pmatrix} \kappa -1 \\ j \end{pmatrix}$ choices to distribute these $j$ vertices in this vertex' 1-neighbourhood remain. The last factor gives the number of ways how the other $I-j$ vertices can be distributed on the remaining $N-(\kappa-1)$ holes of the pattern.}
\bibitem{FN1} {This can be understood as a result of two competing mechanisms. On the one hand, higher $t_{\text{u}}$ demands less vertices to be taken out in order to make a hole unstable. On the other hand, however, the effort to remove a vertex grows with $t_{\text{u}}$. It can be easily verified that the second tendency outweighs the first one.}
\end{thebibliography}
\end{document}